\documentclass[usenatbib]{mnras}  

\usepackage{newtxtext,newtxmath}

\usepackage[T1]{fontenc}

\DeclareRobustCommand{\VAN}[3]{#2}
\let\VANthebibliography\thebibliography
\def\thebibliography{\DeclareRobustCommand{\VAN}[3]{##3}\VANthebibliography}

\usepackage{subcaption} 
\usepackage{rotating}   
\usepackage{graphicx}	
\usepackage{amsmath}	
\usepackage{amssymb}	
\usepackage{lscape}     
\usepackage{cases}      
\usepackage{multirow}   
\usepackage{tikz}       
\usepackage{overpic}    
\usepackage{anyfontsize}
\usepackage[normalem]{ulem}
\defcitealias{Rey2025}{Paper I}
\newcommand{\ARCHITECT }{\textsc{architect}}
\newcommand{\KI}{\texttt{ME}}
\newcommand{\KR}{\texttt{MT}}
\newcommand{\DC}{\texttt{DC}}
\newcommand{\HI}{\ion{H}{i}}
\newcommand{\HeI}{\ion{He}{i}}
\newcommand{\HeII}{\ion{He}{ii}}
\newcommand{\MgII}{\ion{Mg}{ii}}
\newcommand{\CIV}{\ion{C}{iv}}
\newcommand{\OVI}{\ion{O}{vi}}

\title[Impact of subgrid physics on the CGM]{ARCHITECTS II: Impact of subgrid physics on the observable properties of the circumgalactic medium}

\author[M. Rey et al.]{Maxime Rey$^{1,2}$\thanks{E-mail: maximerey.astro@gmail.com},
Jérémy Blaizot$^{2}$,
Taysun Kimm$^{1}$,
Joakim Rosdahl$^{2}$,
Léo Michel-Dansac$^{3}$,
Valentin Mauerhofer$^{4}$
\\
$^{1}$Department of Astronomy, Yonsei University, 50 Yonsei-ro, Seodaemun-gu, Seoul 03722, Republic of Korea \\
$^{2}$Centre de Recherche Astrophysique de Lyon UMR5574, ENS de Lyon, Univ. Lyon, CNRS, 9 av. Charles André, F-69230 Saint-Genis-Laval, France \\
$^{3}$Aix Marseille Univ., CNRS, CNES, LAM, Marseille, France\\
$^{4}$Kapteyn Astronomical Institute, University of Groningen, PO Box 800, 9700 AV Groningen, The Netherlands\\
}

\date{Accepted XXX. Received YYY; in original form ZZZ}

\pubyear{2022}

\begin{document}
\label{firstpage}
\pagerange{\pageref{firstpage}--\pageref{lastpage}}
\maketitle

\begin{abstract}
Galaxy evolution is driven by star formation and stellar feedback on scales unresolved by current high-resolution cosmological simulations, requiring robust subgrid models. However, these models remain degenerate, often calibrated primarily to match observed stellar masses.
To explore these degeneracies, we conduct three state-of-the-art cosmological zoom-in simulations of the same galaxy, each incorporating different subgrid models: mechanical feedback, a combination of mechanical and thermal feedback, and delayed cooling. We compare their circumgalactic media (CGM) through quasar absorption sightlines of \HI, \MgII, \CIV, and \OVI. 
Our findings demonstrate that despite producing galaxies with the same stellar masses, the models lead to distinct feedback modes and CGM properties. Column densities and covering fractions serve as effective diagnostics of subgrid models, with all four ions providing strong constraints as they trace diverse gas phases, exhibit complementary spatial distributions, and originate from different mechanisms. 
Although all simulations bracket observed column density distributions, direct comparisons are limited by scarce detections and significant scatter in absorption strengths. Covering fractions of weak absorbers provides the most robust constraints.
All models fail to reproduce \HI\ and \MgII\ covering fractions, and delayed cooling overproduces \OVI\ covering fractions, while the other models underproduce them. The simulation including mechanical feedback reproduces the observed \CIV\ covering fractions well, whereas the other models show slight offsets. We argue that this discrepancy is likely driven by unresolved thermal structures for \HI\ and \MgII\, and insufficient metals for \CIV\ and \OVI, arising from missing physics such as AGNs or cosmic rays.
\end{abstract}



\section{Introduction}
The circum-galactic medium (CGM) plays a pivotal role in the formation, growth, and evolution of its host galaxy \citep{Tumlinson2017, Faucher-Giguere2023, Peroux2024}. As matter falls from cosmic filaments onto the interstellar medium (ISM) to fuel galaxy formation \citep{Hafen2019, Afruni2021, Saeedzadeh2023, Decataldo2024}, it passes through the CGM. There, the inflowing gas interacts with hot gas ejected from the galaxy through different feedback mechanisms \citep{Crain2013, Mitchell2018, Peroux2020b}. Additionally, recent studies also suggest the presence of small cold gas clouds entrained along hot outflowing winds \citep{Gronke2018, Gronke2020b, Gronke2023, Tan2024}, making the CGM a highly multiscale multiphase medium \citep{Tripp2008, Chen2009, Prochaska2011, Johnson2015, Zheng2024, Zou2023, Mishra2024}. Understanding how these different phases and flows interact and impact each other is crucial to understand how the baryon cycle operates and how galaxies regulate their growth.

From the observational viewpoint, the CGM is challenging to study due to its diffuse nature. Even though a few studies are starting to observe the CGM in emission \citep{Guo2023, Leclercq2024, Dutta2024, Kusakabe2024}, the main constraints still come from quasar absorption lines, notably thanks to the Cosmic Origins Spectrograph observations (COS-Dwarfs in \citealt{Bordoloi2014}, COS-Halos in \citealt{Prochaska2017}, COS-legacy/HSLA in \citealt{Peeples2017}, CGM$^2$ in \citealt{Wilde2021}, COS-Holes in \citealt{Garza2024}), but also more recently with new instruments used for the Dark Energy Spectroscopic Instrument (DESI) and Cosmological Evolution Survey (COSMOS) surveys \citep{Zou2023}. To probe its multiphase nature, different ions are studied, typically low-ions such as \HI\ \citep{Thom2012, Prochaska2017, Wilde2021, Wilde2023, Dutta2024b, Augustin2024}, \MgII\ \citep{Chen2010, Nielsen2013, Lan2020, Huang2021}, \ion{C}{ii} and \ion{Si}{ii}\ \citep{Werk2013} for cold gas, \ion{Si}{iii-iv} \citep{Zheng2024} or \CIV\ \citep{Bordoloi2014, Borthakur2013, Manuwal2021} for warm gas, and \OVI\ \citep{Tumlinson2011, Thom2008}, \ion{O}{viii}\ \citep{Locatelli2024} or \ion{Ne}{viii} \citep{Qu2024} for hotter phases of the gas. Quasar absorption lines provide discrete probes of the CGM for individual galaxies. Consequently, it is necessary to combine observations across large galaxy samples to reach a better understanding of the physical state of the CGM. Current constraints are limited by sample size, making careful selection critical, as CGM properties could depend on host galaxy properties \citep{Anand2021, Dutta2021, Zou2023, Wu2025, Chen2025}. Furthermore, the detection limits of shallow surveys may introduce bias by failing to capture weak absorption features.

From the theoretical viewpoint, the CGM also remains a major challenge for galaxy formation simulations, as it is necessary to properly capture both the cosmological context driving inflows from cosmic filaments and small-scale stellar physics triggering outflows. The zoom-in strategy allows theorists to produce megaparsecs-scale simulations with resolutions of a few hundred parsecs in the CGM \citep{Sawala2016, Fattahi2016, Grand2017, Dubois2021}. Yet, the resolution in most numerical simulations is focused on the densest regions in the ISM, while the diffuse CGM remains poorly resolved, far above kiloparsec scales. Recent studies find the CGM to consist of a hot corona populated by cold dense clouds with sizes smaller than a hundred parsec \citep{Hennawi2015, Lau2016, McCourt2018, Faucher-Giguere2023}. These clouds are typically unresolved in such simulations and disappear as they spuriously mix with hot ambient gas. Simulations thus often struggle to simultaneously reproduce both high and low ion constraints \citep{Hummels2013, Ford2016, Liang2016, Suresh2017, Gutcke2017, Appleby2021, Mallik2023, Zheng2024}, and they may have to use subgrid models for the CGM \citep[e.g.][]{Nikolis2024, Hummels2024, Butsky2024, Das2025}.

In parallel, a significant challenge in simulations of galaxy formation is to discriminate between the various subgrid models used to describe star formation and feedback processes. Traditionally, comparative studies focus on the impact of these models on the global properties of galaxies \citep{Kimm2015, Nunez-Castineyra2021, Azartash-Namin2024, Kang2025}. However, many subgrid models can be tuned to match galactic observational constraints \citep{Pillepich2018, Tremmel2017, Lee2021, Schaye2023, Kim2025}, leaving them degenerate with no consensus on a favoured model. To break this degeneracy, additional observables must be considered. With its complex and dynamic nature, particularly challenging to replicate in simulations, the CGM is a promising candidate to discriminate between different feedback models while learning more about the processes regulating galaxy formation. Recognising this potential, several studies have begun to leverage CGM properties as a diagnostic tool to explore the effects of various feedback mechanisms and model choices. These include investigations into supernovae \citep{Hummels2013, Ford2016, Liang2016}, cosmic rays \citep{Salem2016, Ji2020, DeFelippis2024}, active galactic nuclei (AGN) feedback \citep{Sanchez2019, Sanchez2024, Ramesh2023}, AGN flickering \citep{Segers2017, Oppenheimer2018}, the UVB \citep{Corlies2016, Hussain2017, Mallik2023, Taira2025}, and even different refinement strategies in simulations \citep{Hummels2019, Peeples2019, Suresh2019, vandeVoort2019, Ramesh2024b}. 
However, when comparing these feedback implementations, the mass of the galaxies simulated can be significantly affected. As CGM properties are found to depend on star-forming properties \citep{Lan2014, Zou2023, Appleby2023, Wu2025, Chen2025} and galaxy mass in both observations \citep{Bordoloi2011, Menard2011, Bordoloi2018, Anand2021, Dutta2021} and galaxy simulations \citep{Nelson2020, Ramesh2023, Cook2025}, such comparisons do not isolate the effect of feedback itself. Hence, \citet{Rey2025} (henceforth \citetalias{Rey2025}) carried out the \ARCHITECT\ suite of zoom-in simulations designed to isolate the impact of three different subgrid models commonly used in simulations. They did so by calibrating the stellar mass of the three high-resolution simulations and showed that even with similar stellar masses and star formation rates (albeit different burstiness), the outflow properties are distinct in the three models. Thus, the baryonic content, metallicity, and temperature of the CGM are all also greatly affected by the subgrid model employed. While \citetalias{Rey2025} focused on how the physical properties of the CGM vary with subgrid models, the present follow-up paper proposes a detailed comparison of the simulations with observations, and discusses the physical origin of successes and failures in this exercise.

In this second paper from the \ARCHITECT\ series, we first summarise the properties of the \ARCHITECT\ simulations in Section \ref{sec:methods} before detailing the computation and properties of the four ions (\HI, \MgII, \CIV\ and \OVI) examined in Section \ref{sec:ioninCGM}. With these simulations, we evaluate the sensitivity of CGM mock observations to diverse subgrid prescriptions and determine whether such constraints can effectively discriminate them. Concurrently, we compare the state of the modelled CGM to observations. We finally discuss the limits of observations and of our simulations in Section \ref{sec:discussion}, before concluding in Section \ref{sec:conclusions}.

\section{Numerical methods} \label{sec:methods}
    The setup of the \ARCHITECT\ simulations is fully detailed in \citetalias{Rey2025}. In this section, we briefly summarise key elements of the setup and the main characteristics and differences of the three models. 

    \subsection{Numerical setup}
        The \ARCHITECT\ simulations are run with the hydrodynamical adaptive mesh refinement code RAMSES \citep{Teyssier2002, Teyssier2010}. This code solves the Euler equations of hydrodynamics, including self-gravity and radiative cooling and heating, and incorporates non-equilibrium hydrogen and helium photochemistry coupled to on-the-fly radiative transfer \citep{Rosdahl2013, Rosdahl2015b}. We model dark matter and stellar populations as collisionless particles with minimum masses of $m_{\rm DM} = 3.49 \times 10^5\rm\,M_\odot$ and $m_\star = 3.2 \times 10^3\ \rm\,M_\odot$, respectively. The particles only interact gravitationally using a particle-mesh solver and cloud-in-cell interpolation \citep{Guillet2011}. The cosmological ICs are generated with MUSIC \citep{Hahn2011, Hahn2013}, using cosmological parameters $\Omega_\Lambda = 0.6825$, $\Omega_\mathrm{m} = 0.3175$, and $\Omega_\mathrm{b} = 0.049$ \citep{PlanckCollaboration2014}. 

        We use the Binary Population and Spectral Synthesis model \citep[v2.2.1, ][]{Stanway2018} to model stellar spectral energy distributions. We do not account for non-ionising radiation and follow three photons groups defined by the ionisation energies of \HI, \HeI, and \HeII\ (tracked in the simulation), with energies in eV of [13.6, 24.59), [24.59, 54.42), [54.42, $\infty$), for which we model photoionisation, cooling and heating, and momentum transfer \citep{Rosdahl2015b}. Radiation is transported through the grid at a reduced speed of light of $\mathrm{c_{sim}} = 0.005\rm\,c$ \citep{Gnedin2001, Rosdahl2013}. We also turn on a uniform UV background at $z=8.5$ following \citet{Faucher-Giguere2009}, and use an exponential damping factor to model self-shielding above gas densities of $n_\mathrm{H}=10^{-2}\rm\,cm^{-3}$ \citep{Schaye2001, Rosdahl2012b}.
        Metals are tracked through a single variable, and we model their contribution to metal cooling using tabulated results from CLOUDY (\citealp{Ferland1998}, version 6.02) assuming photoionisation equilibrium with the \citet{Haardt1996} UV background above $10^4\rm\,K$, and fine structure metal cooling rates from \citet{Rosen1995} from $10^4\rm\,K$ down to $10\rm\,K$.

        The three simulations follow the same $3\,R_\mathrm{vir}$ zoom-in region within a $30\rm\,cMpc\,h^{-1}$ box, focusing on a relatively isolated $M_{\rm h} = 5.3 \times 10^{11}\rm\,M_\odot$ mass halo at $z=0$ with a minimum cell width of $\approx 40\rm\,pc$. 
        We adaptively refine a cell, i.e. split it into 8 children cells, when {$M_\mathrm{DM, cell}+\frac{\Omega_m}{\rm\Omega_b} M_\mathrm{b,cell}>8m_\mathrm{DM}$, with $M_\mathrm{DM, cell}$ and $M_\mathrm{b,cell}$ the dark matter and baryonic mass contained within a cell, or when the local thermal Jeans length is smaller than four cell widths.

        The three \ARCHITECT\ simulations differ in their subgrid models for star formation and stellar feedback. We calibrate the subgrid models such that the three galaxies reach a comparable stellar mass at $z=1$. However, note that photo-ionisation, heating and radiation pressure from young stars are included in the same way in all three models, as it is accounted for in the RHD solver.

        \vspace{0.3 cm} 
        The first simulation, which we refer to as \KI\ uses a thermo-turbulent subgrid model for star formation \citep{Kimm2017}, and mechanical supernova feedback by \citet{Kimm2015}. Numerous RAMSES simulations rely on these prescriptions, such as SPHINX \citep{Rosdahl2018, Katz2021, Rosdahl2022}, NewHorizon \citep{Dubois2021}, NewCluster \citep{Han2026} and OBELISK \citep{Trebitsch2021}.
        If a gas cell has a density higher than $10\rm\,cm^{-3}$, is a local density maximum, has a converging flow, and a turbulent Jeans length shorter than four cell widths, stellar particles can be formed with an efficiency $\epsilon_\mathrm{ff}$ given by the multi-freefall model introduced by \citet{Hennebelle2011} and \citet{Federrath2012} (see \citetalias{Rey2025} for more details). The mass of the stellar particles created is an integer multiple of the stellar mass resolution of $\approx 3200\rm\,M_\odot$, though it must be lower than $90\%$ of the cell gas mass. The particle mass is Poisson sampled \citep{Rasera2006} from the mean stellar mass formed within a timestep $\Delta t$: $\overline{m}_\star = \rho \Delta x^3 \frac{\Delta t}{t_\mathrm{ff}} $, with $\rho$ and $\Delta x$ the density and size of the cell, $t_\mathrm{ff} = [{3 \pi}/{(32 G \rho)}]^{1 / 2}$ the freefall time, and $G$ the gravitational constant.
        Each stellar particle represents a simple stellar population which undergoes type II supernova events $10\rm\ Myr$ after being formed. Type II supernovae are modelled with mechanical feedback, in which momentum and energy are injected into the cells neighbouring the supernova host cell, following the expected evolutionary phase of the supernova (radiative or adiabatic) inferred from the local gas conditions. Supernovae inject metals with a yield of 7.5\%. The criterion to distinguish the two phases is based on the mass swept up by the supernova. To calibrate the models in stellar mass, we increase the supernova rate by a factor of two, such that they explode with a rate of 4 supernovae by $100\,M_\odot$.

        \vspace{0.3 cm} 
        The second model, which we refer to as \KR, is described in \citet{Kretschmer2020}. It is used in the MIGA simulations \citep{Kretschmer2022}, the EMP-Pathfinder suite \citep{Reina-Campos2022} and the SPICE suite \citep{Bhagwat2024}. For star formation, if a cell's density is higher than $0.1\rm\,cm^{-3}$ (in practice, no stars form below densities of $10\rm\,cm^{-3}$), stars also form with an efficiency given by the multi-freefall model, but with on-the-fly subgrid turbulence modelling, a critical density above which the star formation efficiency is computed from \citet{Krumholz2005} instead of \citet{Hennebelle2011}, and different values for the fraction of gas blown away by protostellar feedback and the estimated error on the timescale for gas to become unstable.
        As mentioned previously, we use the same radiation feedback model for all simulations, as it is already modelled by the RHD solver \citep{Rosdahl2013, Rosdahl2015b}. To avoid double-counting, we thus turn off the radiation feedback included in the original model of \citet{Kretschmer2020}. In \KR, supernova explosions are uniformly sampled in time from $3\rm\,Myr$ to $20\rm\,Myr$. The supernova energy and momentum are also injected depending on the phase expected, but with a criterion based on the cooling radius of the host cell \citep{Martizzi2015}. Thermal energy is injected in both cases, but if the cooling losses are important, momentum corresponding to the radiative phase is additionally injected. This momentum is also treated as a source term at the solver level, following \citet{Agertz2013}. Lastly, this model uses a metal yield of 10\%, resulting in a higher metallicity in the galaxy \citepalias{Rey2025}. The supernova rate is increased by a factor of four (i.e. 8 supernovae by $100\,M_\odot$) to calibrate the stellar mass of the galaxy.

        \vspace{0.3 cm} 
        The third and last model, \DC, follows the same star formation recipe as \KI, but the delayed cooling model from \citet{Teyssier2013}. This prescription, similar to the NIHAO model \citep{Wang2015}, is used in the MIRAGE simulations \citep{Perret2014} and commonly employed as a reference for studies comparing feedback models \citep{Dale2015, Gabor2016, Rosdahl2017, Nunez-Castineyra2021} as it tends to regulate star formation efficiently. In this model, the supernova energy is injected as thermal energy $10\rm\,Myr$ after the stellar particle birth. To avoid numerical overcooling, cooling is locally turned off following a passive non-thermal energy density variable advected with the gas, increasing at each supernova explosion, and exponentially damped with a characteristic dissipation timescale of $10\rm\,Myr$ \citep{Teyssier2013}.

\section{Ions in the CGM} \label{sec:ioninCGM}
    In this paper, we aim to compare the properties of the \ARCHITECT\ simulations to CGM observables. We define the CGM as all gas located within the virial radius of the halo, $R_{200}$ — the radius within which the mean density of the halo is 200 times the critical density of the Universe — and excluding gas within $0.1\,R_{200}$, which we categorise as the ISM, following \citetalias{Rey2025}. Over the redshift range $z=1-1.3$, $R_{200}\approx100\,\rm\,kpc$. To compare the \ARCHITECT\ simulations to observations, we choose to focus on four ions: \HI, \MgII, \CIV, and \OVI. These ions are the most commonly used in observational studies and trace a wide range of temperatures, owing to their distinct ionisation potentials: $13.6\rm\,eV$ (\HI), $15\rm\,eV$ (\MgII), $64.49\rm\,eV$ (\CIV), and $138.12\rm\,eV$ (\OVI). \HI\ is complementary to \MgII\ as they trace similar gas phases, but while \MgII\ depends on metallicity, \HI\ does not. In this section, we start by comparing the properties of these four ions.

    \subsection{Properties of the observable gas} \label{subsec:prop_ions}
        The \HI\ content of our simulations is a direct prediction from the \ARCHITECT\ radiation hydrodynamic simulations, and \HI\ is thus the most robust comparison point. For the other species, we assume universal Solar abundance values of $A_\mathrm{Mg, \odot} = 3.98\times10^{-5}$, $A_\mathrm{C, \odot} = 2.69\times10^{-4}$ and $A_\mathrm{O, \odot} = 4.9\times10^{-4}$ \citep{Asplund2009} to derive the number density of Mg, C, and O from the metal distribution of the simulation. We then post-process the simulation outputs with KROME \citep{Grassi2014} to obtain the ionisation fractions of these elements. We do so following \citet{Mauerhofer2021b} and solve for the equilibrium ionisation balance for the chosen elements, relying on each cell's metallicity, density, temperature and ionising radiation flux.
        In addition to the three ionising radiation bins from the simulation, \citet{Mauerhofer2021b} estimate the flux of non-HI-ionising stellar radiation within [$6\rm\,eV,13.6\rm\,eV$) based on the UV background model from \citet{Haardt2012}, with a minimum of 10 more photons per unit time than in the first ionising bin (roughly the flux ratio between the subionising bin and the first ionising bin of the intrinsic SED). We expect this assumption to be unreliable at the CGM scale as ISM gas will preferentially absorb ionising photons over subionising photons. Instead, we estimate the ratio of these two stellar bins based on the UVB. In practice, we integrate the flux from the UV background model by \citet{Faucher-Giguere2009} over both the non-ionising bin ($F_{6-13.6}$) and our first ionising bin ($F_{13.6-24.59}$), and compute their ratio as $f_\mathrm{UVB} = F_{6-13.6} / F_{13.6-24.59}$. Our estimate of the sub-ionising flux is then given by $f_\mathrm{UVB}*F_\mathrm{bin1}$, with $F_\mathrm{bin1}$ the cell's flux of the first ionising bin. Estimating this bin is important to study \MgII\ properties as the photons it contains can ionise \ion{Mg}{i}.
        Furthermore, the UVB contribution for the ionising bins was not included in \citet{Mauerhofer2021b} as they focused on the ISM. However, the UVB radiation field can dominate the radiation field of the CGM at our redshift \citep{Holguin2024} and impact measured column densities \citep{Mallik2023, Taira2025}. We include the UVB in our computation through its photoionisation rates for each element. To do so, we integrate the cross section $\sigma_\mathrm{ion}$ \citep{Verner1996} of every element studied over each radiation bin alongside the FG09 UVB flux, $F_\mathrm{UVB}(\lambda)$, as $\int \frac{F_\mathrm{UVB}(\lambda)}{\mathrm{hc}}\sigma_\mathrm{ion}(\lambda)\lambda\mathrm{d}\lambda$, where $\rm h$ is the Planck constant, $\rm c$ the speed of light, and $\lambda$ the wavelength.
        We model dust absorption by exponentially damping the UVB flux in each cell by its optical depth $\tau_\mathrm{dust} = n_\mathrm{dust} \sigma_\mathrm{SMC} \Delta x$, with ${n_\mathrm{dust} = \frac{Z}{Z_0}\left(n_\HI + 0.01n_{\mathrm{H}\,\textsc{ii}}\right)}$ \citep{Michel-Dansac2020b}. $Z$, $n_\HI$, $n_\HI$, and $\Delta x$ are the metallicity, \HI\ and \ion{H}{ii} densities, and the size of the cell. $Z_0=0.005$ and $\sigma_\mathrm{SMC}\approx3.8\times10^{-22}\rm\,cm^2$ are the mean metallicity in the SMC and its dust cross section averaged between $6\rm\,eV$ and $13.6\rm\,eV$ \citep{Gnedin2008}.
        Finally, we model self-shielding by exponentially damping the UVB ionising flux if the cell's hydrogen density is higher than $10^{-2}\rm\,cm^{-3}$, as in the simulation.

        \begin{figure}
            \centering
            \begin{subfigure}[b]{\columnwidth}
                \includegraphics[trim=0 53 0 0, clip, width=\textwidth]{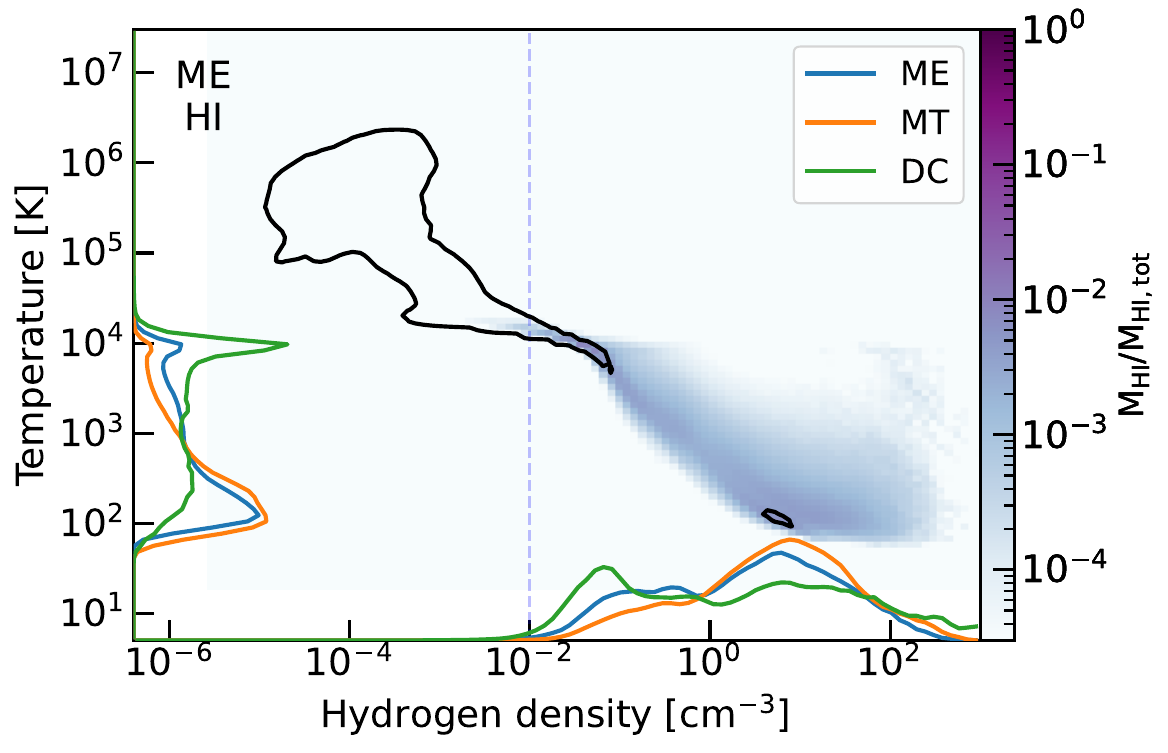}
            \end{subfigure}%
            \hfill
            \begin{subfigure}[b]{\columnwidth}
                \includegraphics[trim=0 53 0 0, clip, width=\textwidth]{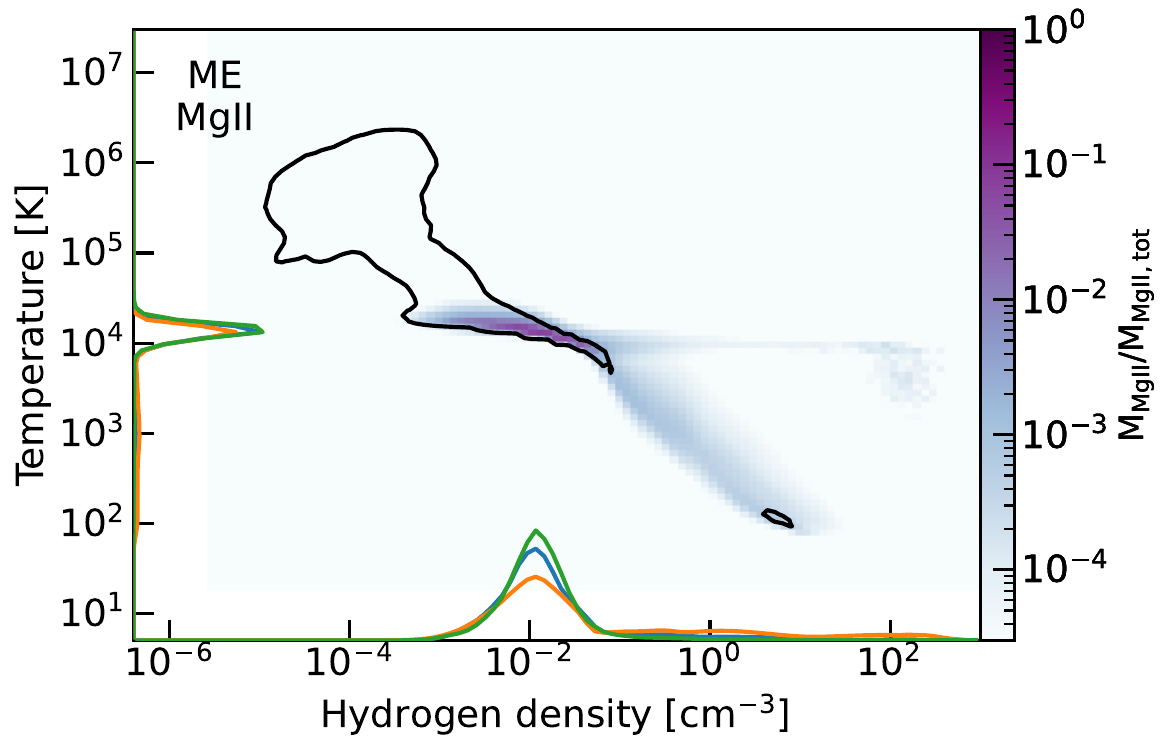}
            \end{subfigure}%
            \hfill
            \begin{subfigure}[b]{\columnwidth}   
                \includegraphics[trim=0 53 0 0, clip, width=\textwidth]{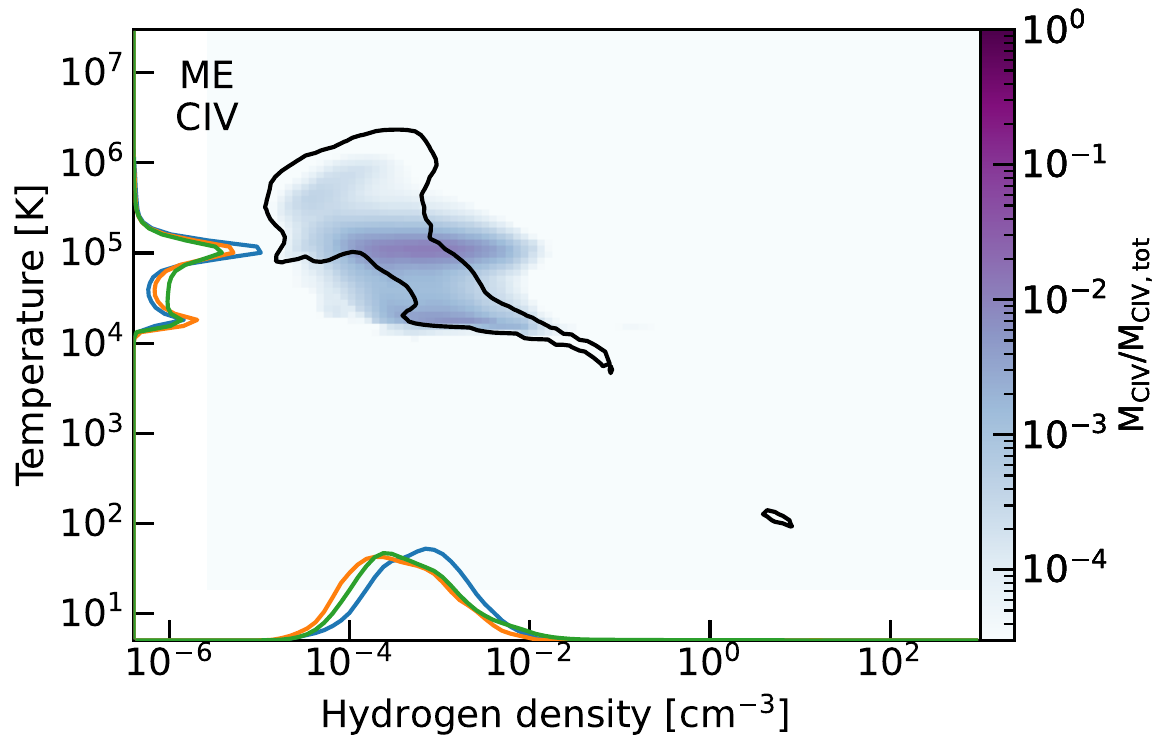}
            \end{subfigure}%
            \hfill
            \begin{subfigure}[b]{\columnwidth}   
                \includegraphics[width=\textwidth]{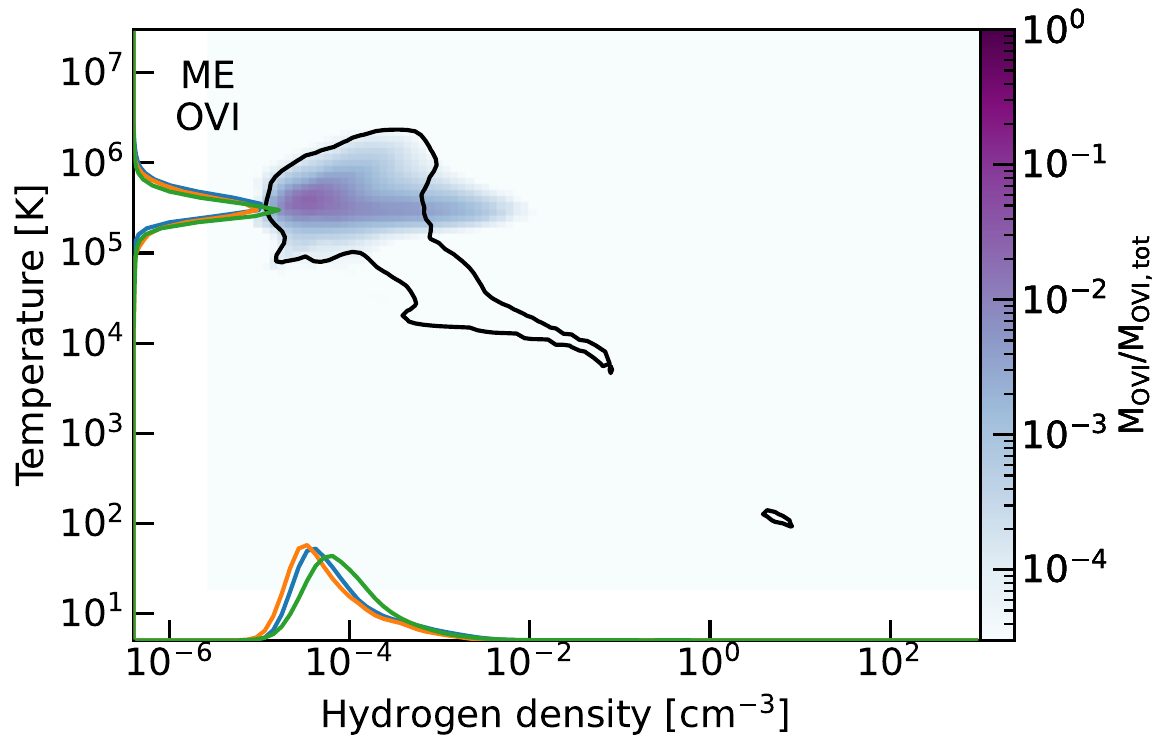}
            \end{subfigure}
            \caption{Phase diagrams of the gas contained in the CGM with \KI. The result consists of 100 snapshots stacked over $1\rm\,Gyr$, from $z=1.3$ to $z=1$. From top to bottom, the phase diagram is weighted by the normalised mass of \HI, \MgII, \CIV\ and \OVI. The solid line contour encompasses 90\% of the total gas mass within the CGM in \KI.
            On the left side and at the bottom of each panel, we show the stacked, normalised mass-weighted PDF of temperature and density in the CGM for the three simulations.
            The vertical blue dashed line in the top panel delimits the densities over which self-shielding is applied on-the-fly with an exponential damping factor. Hydrogen on the left part of this diagram can be photoionised by the UVB.}
            \label{fig:PD_ions}
        \end{figure}

        \vspace{0.3 cm}
        In Fig.~\ref{fig:PD_ions}, we show gas phase diagrams weighted by the mass of the different ions in the CGM with \KI\ over $z=1-1.3$.
        The top panel shows the result for neutral hydrogen. We find that \HI\ traces cold and dense gas at $10^2-10^4\rm\,K$ and hydrogen number densities above $0.01\rm\,cm^{-3}$. Although the black contour indicates that most CGM gas occupies a different region of the phase diagram, hydrogen at such densities is not fully self-shielded and is readily ionised by the local radiation field or the UVB. At densities higher than $100\rm\,cm^{-3}$, neutral hydrogen can turn into molecular hydrogen, which we do not include in our simulation; the corresponding \HI\ mass fractions are therefore likely over-estimated. We find gas at such high densities only within the ISM of the simulated galaxy or within satellites. While \HI\ at these high densities is virtually absent in the CGM, it is made visible here due to the stacking of data over multiple timesteps. Neglecting the formation of H$_2$ thus has little impact on our results regarding the CGM. 
        Focusing on the temperature PDF, we find that while \KI\ and \KR\ exhibit relatively similar distributions with a peak at $10^2\rm\,K$, \DC\ peaks at $10^4\rm\,K$. This difference arises from the efficient nature of the delayed cooling feedback, which creates powerful bursts that disrupt the galaxy and prevent the gas from cooling down to $10^2\rm\,K$. As a result, most of the gas in \DC\ remains at $10^4\rm\,K$, naturally enhancing the \HI\ abundance at these temperatures. \HI\ is also found at similar temperatures in the other two simulations, but in much smaller amounts.\\
        We now focus on \MgII\ in the second panel of Fig.~\ref{fig:PD_ions}. Given the similar ionisation potentials of \HI\ ($13.6\rm\,eV$) and \MgII\ ($15\rm\,eV$), one might expect similar behaviors for these two ions. However, \MgII\ primarily traces gas at densities of $10^{-2}\rm\,cm^{-3}$ and temperatures of $10^4\rm\,K$, while \HI\ is more associated with cooler, denser gas.
        By comparing the phase diagram to the black contour, we find that the \MgII\ peak coincides with a region where most of the CGM gas resides. Owing to its slightly higher ionisation potential, \MgII\ can remain present at higher temperatures before being ionised, unlike \HI, which is ionised at $10^4\rm\,K$. Furthermore, while \MgII\ can recombine to \ion{Mg}{i} at higher densities, \HI\ is already in the ground state, allowing it to trace a broader density range. As a result, \MgII\ traces a similar gas phase in all three models, while \HI\ is more sensitive to the gas distribution. Nonetheless, for \MgII, \DC\ exhibits the highest peak in the mass-weighted PDF, while \KR\ exhibits a wider distribution with a low-density tail extending up to $10^{3}\rm\,cm^{-3}$. The density distribution in \KI\ lies between the two other models.\\
        \CIV\ (third panel) also shows similar distributions across the three models. The \CIV\ density PDF exhibits a broad peak over $\sim10^{-4}-10^{-2}\rm\,cm^{-3}$, while the temperature distribution has two peaks, with the main peak centered at $\sim 10^{5} \rm\,K$, and a second located slightly above $\sim 10^{4} \rm\,K$. 
        While the first peak is located in a region of the phase diagram with little gas \citepalias[see][]{Rey2025}, the \CIV\ ionisation fraction is maximal at these temperatures. Conversely, the second peak arises from a region of the phase diagram where the ionisation fraction is low, but the large reservoir of gas at $10^{4.2} \rm\,K$ (see Fig.~4 in \citetalias{Rey2025}) produces a noticeable peak.\\
        Lastly, \OVI\ (fourth panel) traces hot gas ($10^{5.3}-10^{5.9} \rm\,K$) at low densities ($\sim 10^{-4}\rm\,cm^{-3}$). It preferentially traces the cooler end of the hot halo gas, with its distribution peaking around $\sim 10^{5.5}\rm\,K$ whereas the bulk of the hot gas peaks at higher temperatures, $\sim 10^{5.6}-10^{5.8}\rm\,K$ \citepalias{Rey2025}. O$\,\textsc{vii}$ would provide a better tracer for this hot phase as it peaks at higher temperatures and spans a broader range of ionising temperatures \citep{Tumlinson2017}. However, \OVI\ is rarely observed in the CGM.
        As for \MgII\ and \CIV, \OVI\ shows comparable density and temperature distributions across the three simulations.

        \begin{figure}
            \centering
            \begin{subfigure}[b]{\columnwidth}
                \includegraphics[trim=0 53 0 0, clip, width=\textwidth]{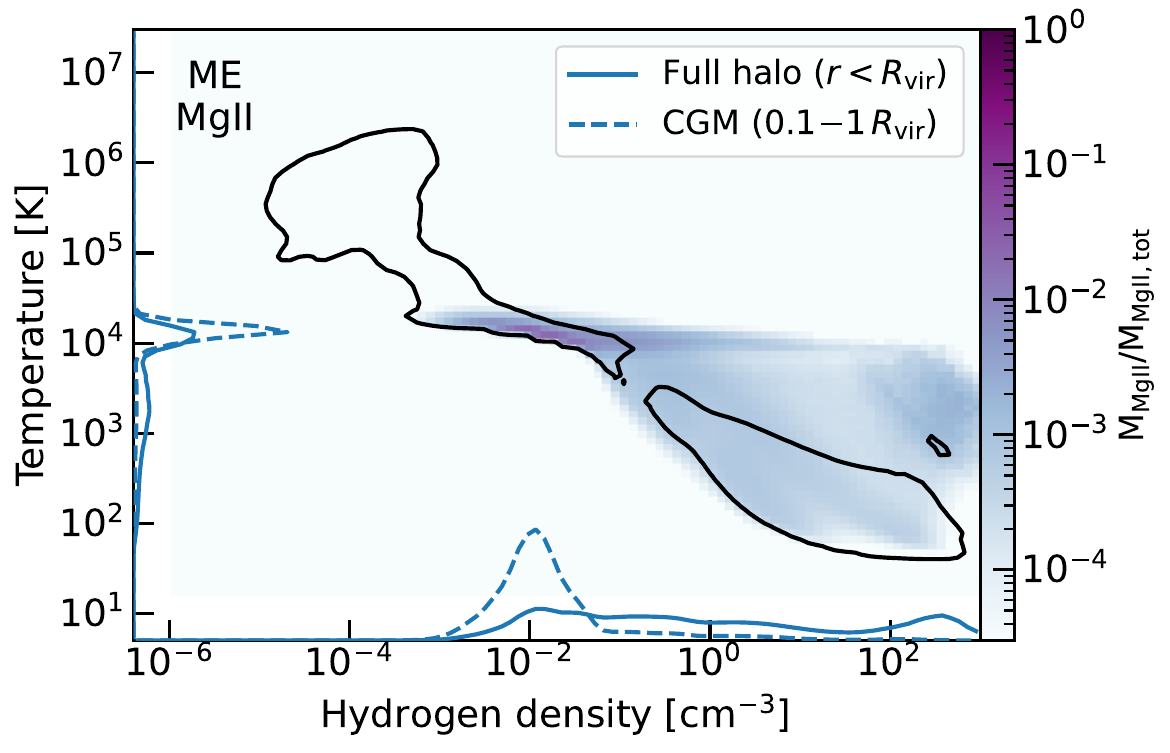}
            \end{subfigure}%
            \hfill
            \begin{subfigure}[b]{\columnwidth}
                \includegraphics[trim=0 53 0 0, clip, width=\textwidth]{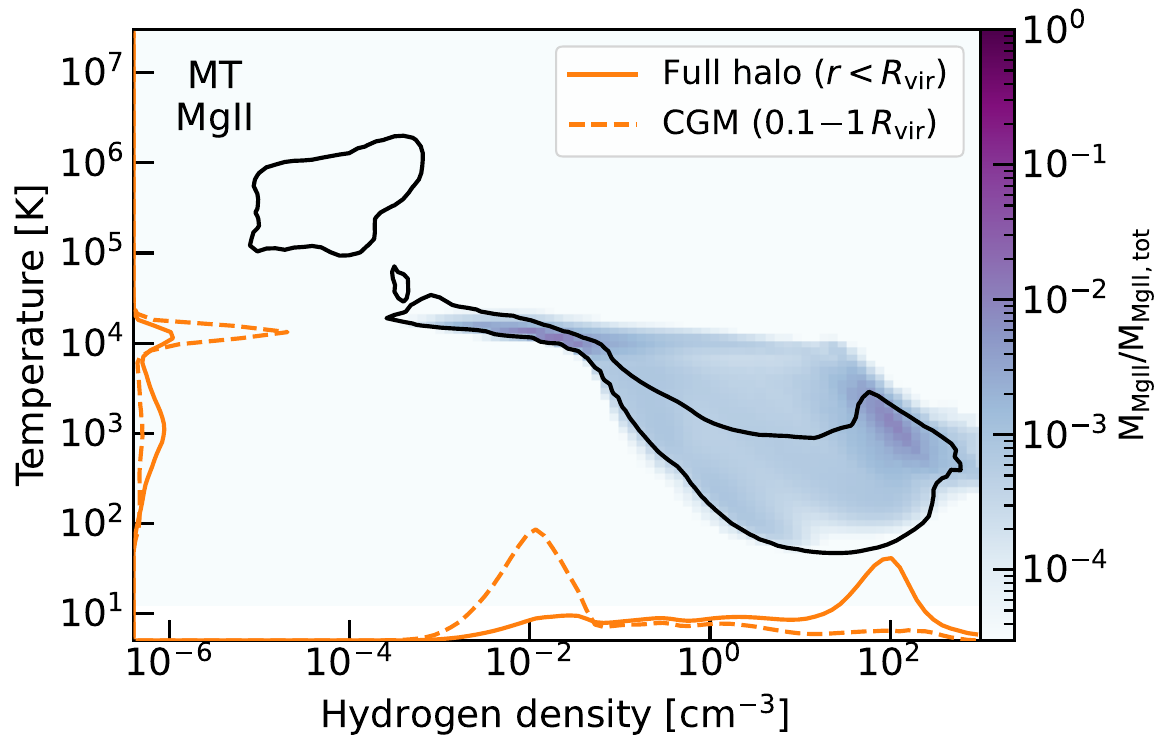}
            \end{subfigure}%
            \hfill
            \begin{subfigure}[b]{\columnwidth}   
                \includegraphics[width=\textwidth]{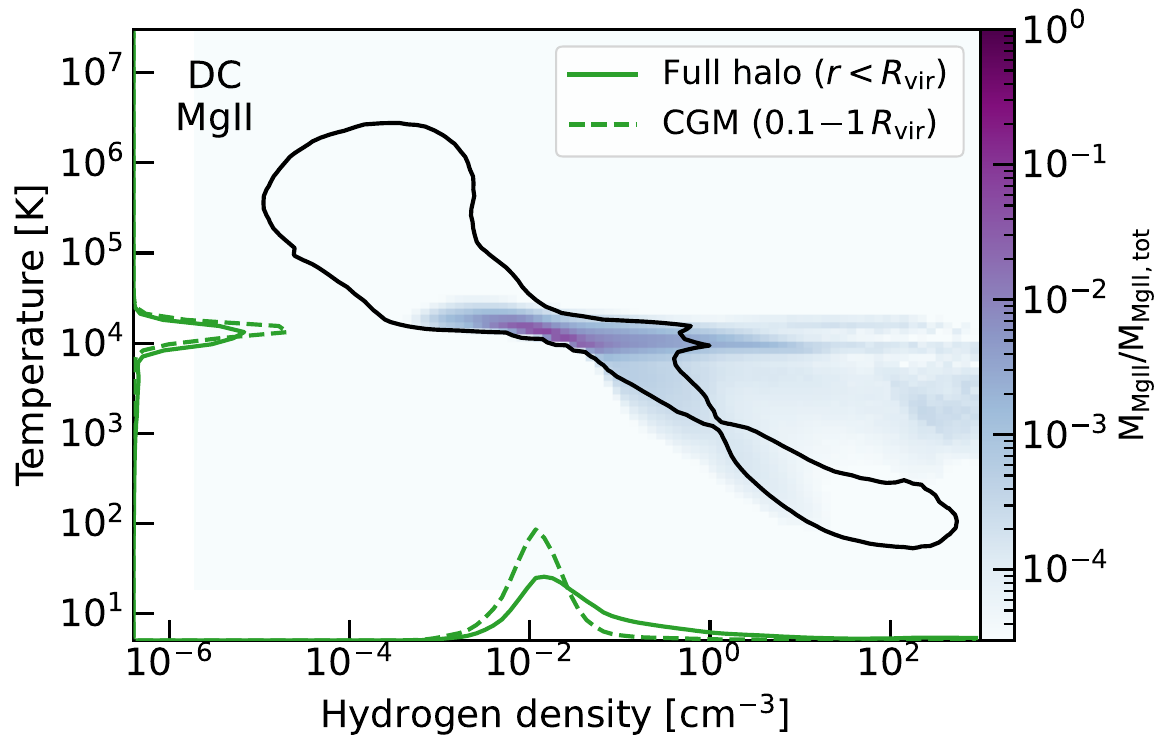}
            \end{subfigure}
            \caption{Phase diagrams of the gas contained within $R_{200}$ in \KI\ (top), \KR\ (top) and \DC\ (bottom), weighted by the normalised mass of \MgII\ in the halo. The result consists of 100 snapshots stacked over $1\rm\,Gyr$, from $z=1.3$ to $z=1$. The solid line contour encompasses 90\% of the full gas mass within $R_{200}$ in each simulation. On the left side and at the bottom of each panel, we show the stacked, normalised, mass-weighted PDF of temperature and density of the corresponding simulation in the whole halo (solid line) and solely in the CGM (dashed line).}
            \label{fig:PD_ions_DC}
        \end{figure}

        \vspace{0.3 cm}
        When comparing \HI\ in the whole halo and in the CGM in \KI\ (not shown), we find very similar distributions, with \HI\ preferentially at $\sim10^2\rm\,K$. For all models, the \CIV\ (resp. \OVI) temperature and density distributions also remain quite similar when comparing in the whole halo and when restricted to the CGM, as most of the warm (resp. hot) gas is already in the CGM. However, the \MgII\ distributions differ significantly when considering either the CGM or the whole halo.
        In Fig.~\ref{fig:PD_ions_DC}, we show the same phase diagrams as in Fig.~\ref{fig:PD_ions}, weighted by \MgII\ mass in the whole halo for the \KI\ (top), \KR\ (middle) and \DC\ (bottom) simulations. We also plot the temperature and density-weighted PDFs, both for the whole halo (solid lines) and the CGM (dashed lines). 
        In \MgII, the density PDF in \KR\ limited to the CGM is in stark contrast with that of the whole halo. The density peak at $10^{-2}\rm\,cm^{-3}$ disappears, while a peak at $10^{2}\rm\,cm^{-3}$ appears. Because the halo in \DC\ contains little gas at $10^{2}\rm\,cm^{-3}$, the density PDF in the CGM of \DC\ is comparable to that of the whole halo, although \MgII\ is even more concentrated in the $10^{-2}\rm\,cm^{-3}$ peak. 
        In temperature, all three simulations exhibit a peak for \MgII\ in the CGM at $10^4\rm\,K$, but their behaviour diverges when considering the whole halo. While \DC\ is only found at $10^4\rm\,K$, \KR\ has a significant fraction of gas around $10^3\rm\,K$. As for the CGM density and temperature distributions, \KI\ stands in between \KR\ and \DC\ when considering the whole halo, with a dominant peak in density at $10^{-2}\rm\,cm^{-3}$ and a tail extending up to $10^{3}\rm\,cm^{-3}$.
        Furthermore, the \MgII\ temperature and density distributions are in differ markedly from those of \HI, showing that despite the similar ionisation potential of \HI\ and \MgII\, they do not trace the same gas phase. While \MgII\ traces $\sim10^4\rm\,K$ gas in the CGM, \HI\ mostly traces $\sim10^2\rm\,K$ gas.

        The observed differences in the density and temperature PDFs in the three models are fully explained by considering the phase diagram of the total gas content. \citetalias{Rey2025} (see their Fig.~4, or the black contours in Fig.~\ref{fig:PD_ions_DC}) indeed demonstrates that while \DC\ only contains a small fraction of cold, dense gas, this phase accounts for roughly half of the halo gas mass in \KI\ and \KR.

    \subsection{Ion distribution in the CGM}
        \begin{figure*}
            \centering
            \huge{\hspace{-0.5cm} \KI\ \hspace{4.5cm} \KR\ \hspace{4.5cm} \DC}
            \begin{subfigure}[b]{0.3145\textwidth}
                \begin{overpic}[width=\textwidth]{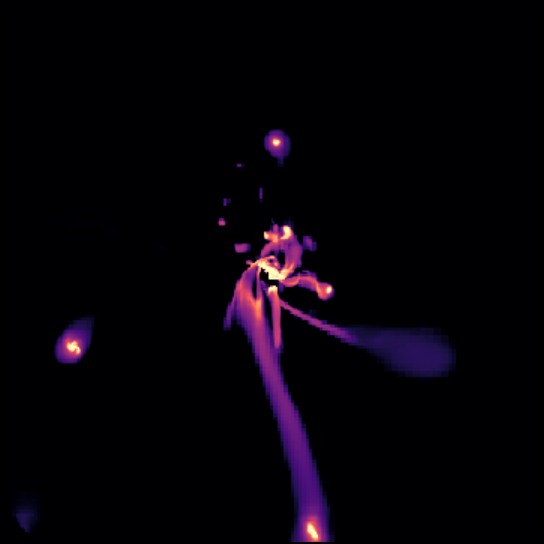}
                    \put(5, 90){\textcolor{white}{\large{HI}}}
                \end{overpic}
            \end{subfigure}%
            \begin{subfigure}[b]{0.3145\textwidth}  
                \includegraphics[width=\textwidth]{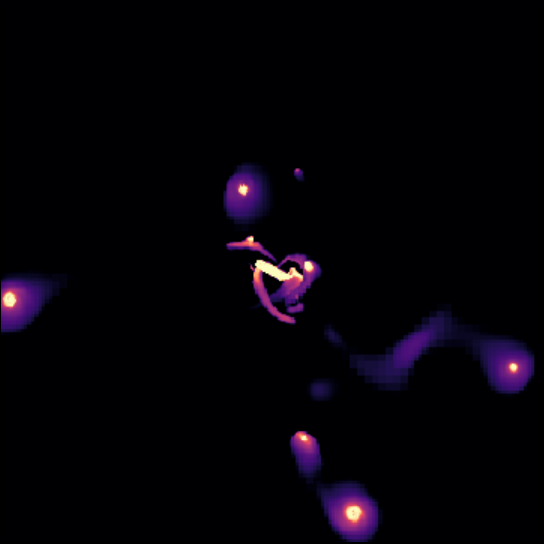}
            \end{subfigure}%
            \begin{subfigure}[b]{0.3395\textwidth}   
                \includegraphics[width=\textwidth]{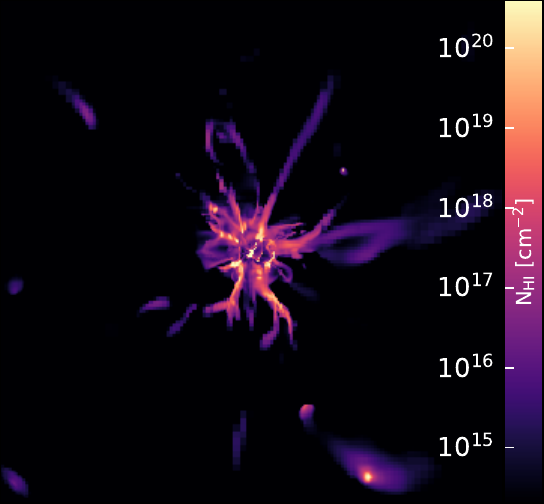}
            \end{subfigure}
            \begin{subfigure}[b]{0.3145\textwidth}
                \begin{overpic}[width=\textwidth]{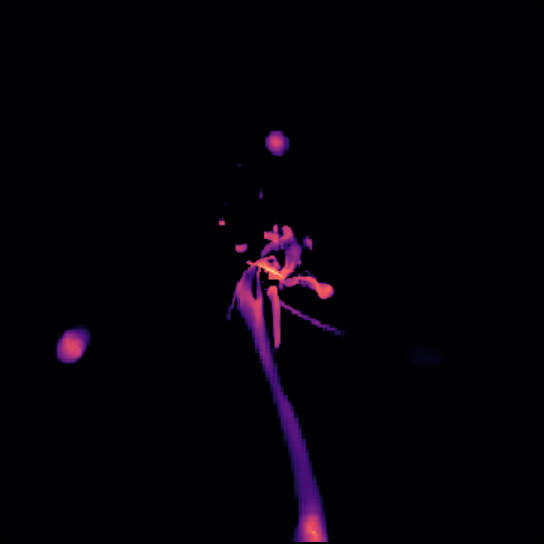}
                    \put(5, 90){\textcolor{white}{\large{MgII}}}
                \end{overpic}
            \end{subfigure}%
            \begin{subfigure}[b]{0.3145\textwidth}  
                \includegraphics[width=\textwidth]{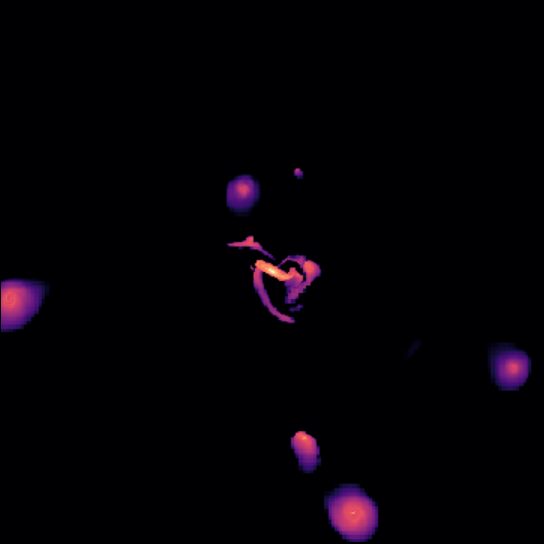}
            \end{subfigure}%
            \begin{subfigure}[b]{0.3395\textwidth}   
                \includegraphics[width=\textwidth]{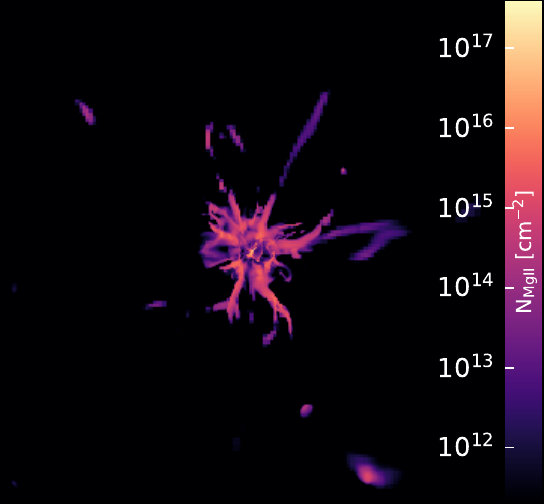}
            \end{subfigure}
            \begin{subfigure}[b]{0.3145\textwidth}
                \begin{overpic}[width=\textwidth]{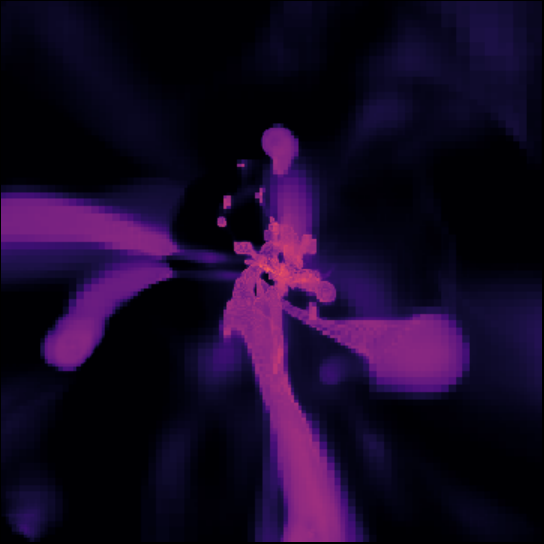}
                    \put(5, 90){\textcolor{white}{\large{CIV}}}
                \end{overpic}
            \end{subfigure}%
            \begin{subfigure}[b]{0.3145\textwidth}  
                \includegraphics[width=\textwidth]{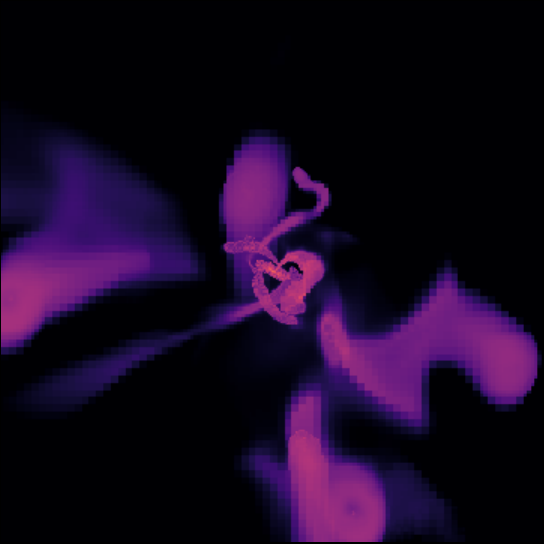}
            \end{subfigure}%
            \begin{subfigure}[b]{0.3395\textwidth}   
                \includegraphics[width=\textwidth]{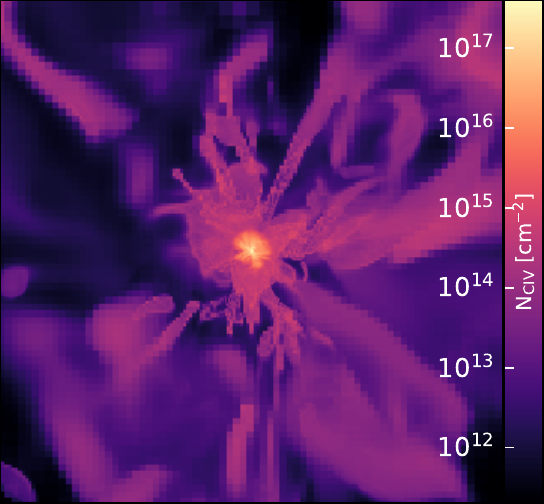}
            \end{subfigure}
    
            \begin{subfigure}[b]{0.3145\textwidth}
                \begin{overpic}[width=\textwidth]{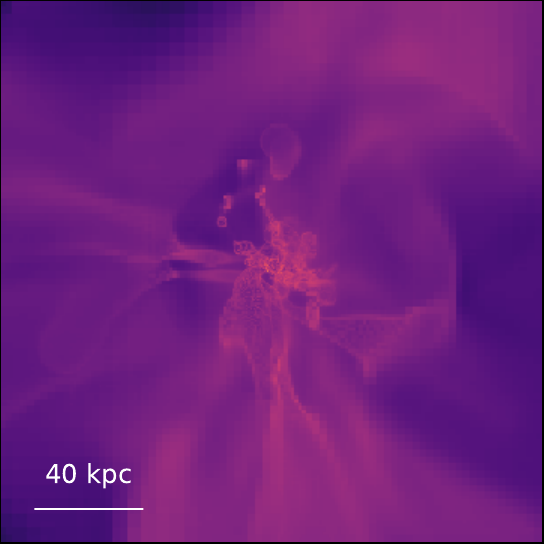}
                    \put(5, 90){\textcolor{white}{\large{OVI}}}
                \end{overpic}
            \end{subfigure}%
            \begin{subfigure}[b]{0.3145\textwidth}  
                \includegraphics[width=\textwidth]{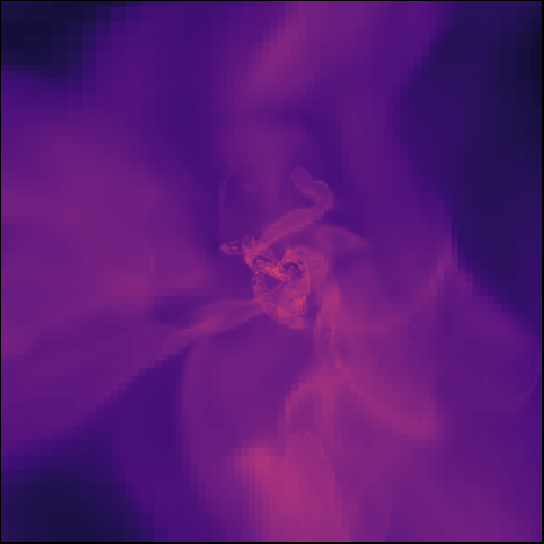}
            \end{subfigure}%
            \begin{subfigure}[b]{0.3395\textwidth}   
                \includegraphics[width=\textwidth]{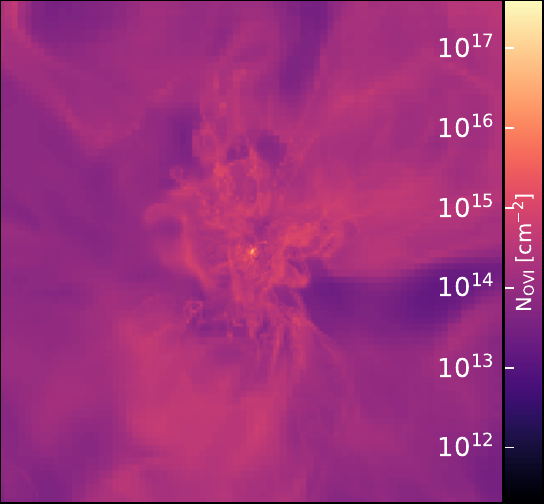}
            \end{subfigure}
            \caption{Maps of the galaxy at $z\approx1.1$ for \KI, \KR, and \DC\ (from left to right). From top to bottom, we show the column density of \HI, \MgII, \CIV\ and \OVI, computed using KROME. The images are $2.5\,R_{200}$ on a side.}
            \label{fig:mapsions}
            \vspace{-5pt} 
        \end{figure*}

        \vspace{0.3 cm}
        To get a clearer image of where the selected ions are in our simulations, we show in Fig.~\ref{fig:mapsions} column density maps of \HI, \MgII, \CIV\ and \OVI\ (from top to bottom) for \KI, \KR, and \DC\ (from left to right). 
        The projection and timestep used for these maps were selected to be representative of the state of the CGM when looking at ion column densities. 
        As expected, \HI\ and \MgII\ trace similar regions, although \HI\ exhibits much higher column densities than \MgII. Both \HI\ and \MgII\ are mostly concentrated in or close to the galaxy and its satellites, and in the filaments of cold gas connecting them. These regions typically have hydrogen densities higher than roughly $\sim 10^{-2}\rm\,cm^{-3}$. \KI\ commonly exhibits dense filamentary structures left by satellites falling towards the central galaxy, while \KR\ has almost no filamentary structures, with \HI\ and \MgII\ forming small haloes around satellite galaxies.
        This discrepancy is likely driven by the differing metal distributions between the models. By comparing density, metallicity, and temperature maps, we find that \MgII\ in \KI\ consists of gas at $n_\mathrm{H}\sim10^{-2}\rm\,cm^{-3}$ and $T\sim10^5\rm\,K$ with metallicities $0.1\,Z_\odot$. In contrast, similar regions in \KR\ have metallicities ten times lower and are hotter and less dense. \citetalias{Rey2025} shows that the CGM metallicity in \KI\ is locally higher than in \KR\ due to different feedback mechanisms. In \KI, feedback ejects sufficient metals from satellites to create high-metallicity trails that cool efficiently, leading to the formation of the \MgII\ structures observed in our maps. Conversely, metals in \KR\ remain largely confined to satellites \citepalias[Fig.~11,][]{Rey2025} and the low-metallicity tails are more readily dissolved into the hot ambient medium. Note, however, that this $10^5\rm\,K$ gas represents only a small fraction of the total \MgII\ in the CGM (see Fig.~\ref{fig:PD_ions}). Except for these tails, \KI\ and \KR\ show close to no \MgII\ at larger distances from the central galaxy or its satellites as gas becomes hotter, less dense, and is ionised by the galaxy or the UVB \citep[eg.][]{Mitchell2021}. In \DC, we find numerous thin, high-density tails stemming from the central galaxy. These are present all around the galaxy, very close to it. Shortly after supernovae explosions, these high \MgII\ column densities trails briefly disappear, hinting at feedback photoionising \MgII\ gas to higher excitation states. Because of the disruptive nature of the delayed cooling feedback, supernova explosions efficiently strip significant chunks of gas from the galaxy, creating high \MgII\ column densities filaments extending further out in the CGM. Overall, the \MgII\ column density distribution in \DC\ evolves very quickly with time but remains high close to the galaxy. In contrast, the distribution of \MgII\ column density in \KI\ and \KR\ evolves at a much lower pace, mainly dictated by satellites' infall.

        Tracing a hotter environment, the spatial distribution of \CIV\ is more extended than \MgII\ or \HI, and shows thicker filaments and bigger haloes around satellite galaxies. \KI\ and \KR\ both exhibit extended filamentary structures in \CIV\ corresponding to high metallicity regions with gas at temperatures at or lower than $\sim10^{5.5}\rm\,K$. By looking at the temporal evolution of \CIV\ column density, we find that it mostly arises from trails left by infalling satellites. Outflows also slightly increase the \CIV\ content in \KI, but only very close to the galaxy. Conversely, \DC\ exhibits significantly different behaviour with \CIV\ column densities notably higher than the other models due to a higher mass of metals in these regions. In contrast to \KI\ and \KR\ where \CIV\ gas is predominantly sourced from satellites, a considerable fraction of \CIV\ in \DC\ comes from outflows. With this model, the dynamics of the \CIV\ gas is much more complicated, with both infalling satellites bringing in enriched gas and large outflowing hot metal-rich plumes ejected beyond the halo scale covering most of the CGM. This results in a significant amount of \CIV\ gas in the CGM of the central galaxy in the \DC\ run, with average \CIV\ column densities of $\sim 10^{14}-10^{15}\rm\,cm^{-2}$. Interestingly, by looking at the time evolution of the \CIV\ density maps, we find that when supernovae explode, they produce a ``flash'' of extremely high \CIV\ column density reaching $\sim 10^{16}-10^{17}\rm\,cm^{-2}$, starting from the galaxy and quickly propagating outwards favouring propagation through high \CIV\ column density channels. We selected in Fig.~\ref{fig:mapsions} a moment when this event begins. The rightmost panel in the third row from the top shows a bright yellow region at the centre of the galaxy, which rapidly propagates outward like the shock front of a supernova. After passing through a given region, the \CIV\ column density quickly falls back to its pre-supernova range. This flash is due to radiation ionising the surrounding gas for a brief amount of time, before it recombines back to its original state.

        Finally, the most energetic ion probed, \OVI, produces similar extended structures as found in \CIV, but also a \textit{diffuse} component permeating the whole CGM with all three models. \OVI\ first appears alongside gas outflows at early times and quickly fills the whole CGM. It is also ejected from infalling galaxies, increasing the complexity of the CGM and bringing in more \OVI. In the bottom rightmost panel of Fig.~\ref{fig:mapsions}, we show that at low redshift, \OVI\ is volume-filling in \DC\ due to multiple episodes of powerful galactic winds. These winds continuously fuel the CGM with \OVI\ and maintain it at high densities in every direction, consistent with the picture of ancient flows producing gas traced by high-energy ions \citep{Ford2014}. The \OVI\ structures in \KR\ more closely resemble accretion flows. These large-scale filaments are mainly high-metallicity tidal tails from flyby satellites, as for \CIV. This nonetheless leads to large regions with low \OVI\ column density as outflows are not powerful enough to maintain the CGM at high \OVI\ column densities. \KI\ stands in between, with a CGM filled with \OVI\ at lower densities than \DC, occasionally replenished by outflows, and filamentary structures from satellites.
        
        Overall, the three ions trace different phases of the CGM, and their spatial distributions are strongly affected by the subgrid modelling of SN feedback. In agreement with \citet{Corlies2016}, we find that the strongest absorbers for low-energy ions tend to reside in the densest structures, mainly galaxies and filaments, whereas high-energy ions are found throughout the volume-filling hot gas.

    \subsection{Column densities and covering fractions} \label{subsec:coldens_measurement}
        \begin{figure}
            \centering
            \begin{subfigure}[b]{0.48\textwidth}
                \includegraphics[width=\textwidth]{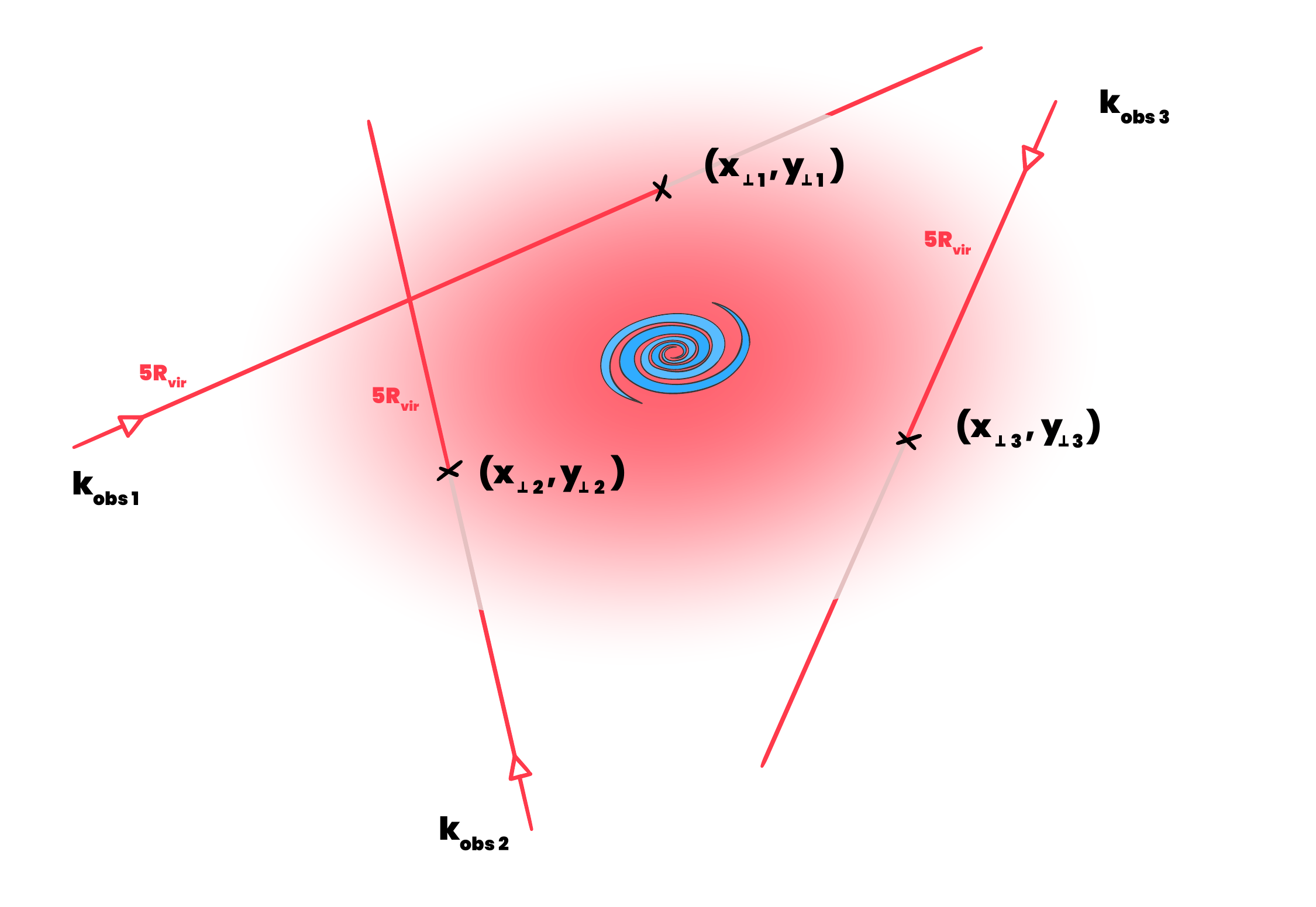}
            \end{subfigure}%
            \caption{Cartoon showing how column densities are obtained in our simulations. The central galaxy is shown in blue, with its CGM represented by the diffuse red halo. We produce $10^5$ rays sampling uniformly impact parameters up to $2\,R_{200}$. Their direction is given by ${\textbf{k}_\mathrm{obs}}$, and they intersect the galaxy plane at a point $(x_{\perp i}, y_{\perp i})$, with $i$ the index of the $i$-th ray.}
            \label{fig:QSO_qbs}
        \end{figure}

        To compare our results with observations, we compute the simulated column densities along random lines of sight by propagating $10^5$ rays in each snapshot with RASCAS \citep{Michel-Dansac2020b}. We illustrate this process in Fig.~\ref{fig:QSO_qbs}, with three different lines of sight. We draw for each ray a random direction $k_\mathrm{obs}$ and two random coordinates, $x_\perp$ and $y_\perp$, on the plane perpendicular to $k_\mathrm{obs}$ and intersecting the halo centre, defined as the x,y origin. We retain only rays with $r_\perp = \sqrt{x_\perp^2+y_\perp^2} \leq 2\,R_{200}$, and thus produce a set of lines of sight sampling uniformly the plane up to impact parameters of $2\,R_{200}$. To obtain column densities, we then integrate the density of ions along these rays over $5\,R_{200}$ ($2.5\,R_{200}$ on each side)\footnote{$5\,R_{200}$ roughly corresponds to $\delta v \sim 5\,R_{200}H(z) \approx \pm 30-35\rm\,km\,s^{-1}$ at $z=1-1.3$.}.

        For the observational data, we select \textit{galaxy-selected} surveys, i.e. surveys which look for absorption lines in quasar spectra for galaxies along or close to their line of sight\footnote{Galaxy-selected surveys are in opposition with absorber-selected surveys, which rely on the detection of absorption features and then try to match galaxies.}. This method is closest to our measurements in the simulation and allows for upper limits on the strength of absorption features. We detail our selection of observational measurements in Appendix \ref{sec:Appendix_obs} and discuss their limits in section \ref{subsec:limits_obs}.

        \subsubsection{Cold gas: H\texorpdfstring{$\,\textsc{i}$}{I} and Mg\texorpdfstring{$\,\textsc{ii}$}{II}}
            \begin{figure}
                \centering
                \includegraphics[width=\columnwidth]{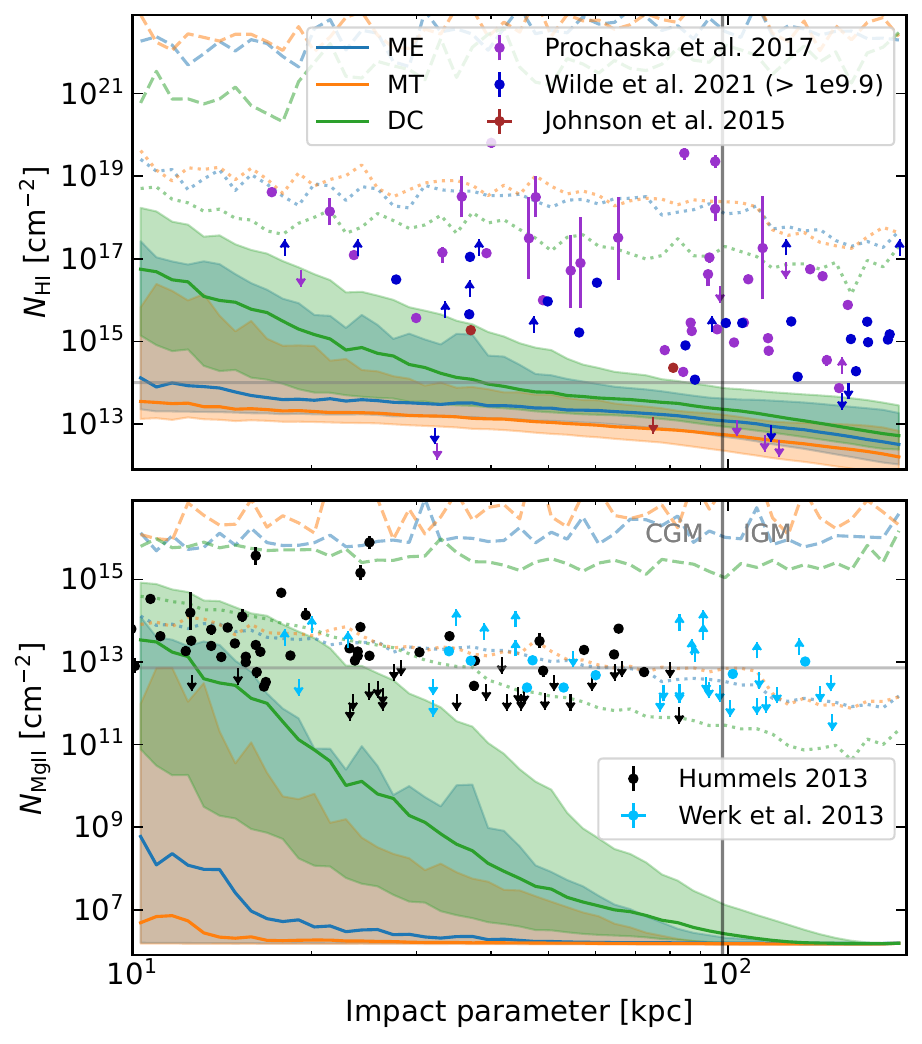}
            \caption{Column density as a function of impact parameter for \HI\ (top) and \MgII\ (bottom) stacked over $1\rm\,Gyr$, from $z=1.3$ to $z=1$. The solid lines correspond to the median column density, and the shaded areas show the 15.9 and 84.1 percentiles. We also show the maximum (resp. mean) column densities measured as dashed (resp. dotted) lines. Different markers show galaxy-selected observational points from different references. Upward-pointing arrows denote lower limits and downward-pointing arrows denote upper limits. The horizontal grey line shows the threshold used for the covering fractions in Fig.~\ref{fig:frac_cold}, and the vertical black line corresponds to the mean $R_{200}$ over our redshift range, $R_{200\rm,avg}\approx98\rm\,kpc$.}
            \label{fig:rad_col_cold}
            \end{figure}
    
            In Fig.~\ref{fig:rad_col_cold}, we show \HI\ and \MgII\ column densities as a function of impact parameter, with the shaded regions representing the 15.9 and 84.1 percentiles. In the upper panel, we see that the column densities of cold gas, traced by \HI, decrease with radius for all simulations. In the inner CGM, the differences between models are the largest, and we see that the \DC\ simulation produces a more extended cold phase, out to $\approx 20\rm\,kpc$, while \KI\ and \KR\ show almost no enhancement of cold gas close to the galaxy. As the feedback is more efficient at ejecting gas in \DC, the cold gas structures of the galaxy are frequently stripped and dragged further than $0.1\,R_{200}$. This leads to lower \HI\ column densities in the ISM and enhanced column densities in the inner CGM (see \citetalias{Rey2025}). In the outer CGM, the column density profiles of the three simulations are very similar, with \DC\ higher than \KI\ and, in turn, \KI\ higher than \KR.
            Moving to the lower panel of Fig.~\ref{fig:rad_col_cold}, we see that the \MgII\ column density profiles show a similar picture, with \MgII\ column densities much higher in \DC\ than with the two other models. We see two possible explanations. We have seen in Fig.~\ref{fig:PD_ions_DC} that \MgII\ preferentially traces $\sim10^4\rm\,K$ gas, which represents a larger mass in \DC\ than in \KR. However, the difference in \MgII\ column density also increases between \DC\ and \KI, which both have a similar cold gas mass in the CGM \citepalias{Rey2025}. As \citetalias{Rey2025} found that the metal mass in the CGM is much larger in \DC\ than in both \KI\ and \KR, we find that this larger difference in \MgII\ column density is driven by metallicity. The \KI\ model remains slightly above the \KR\ model, due to either higher metallicity or colder gas.
            
            Most \HI\ observations are located in the outer regions of the CGM, and most detections find column densities several orders of magnitude above the median column densities from our simulations. In \MgII, the observations probe a larger range of impact parameters, starting at $10\rm\,kpc$. In the inner CGM, \DC\ is in good agreement with \citet{Hummels2013}, while a small fraction of the simulated lines of sight overlap in \KI. Beyond $\approx 30\rm\,kpc$ from the centre, most observations are upper limits plateauing at $\sim 10^{12}\rm\,cm^{-2}$ while simulations show three to four orders of magnitude lower \MgII\ median column densities. Despite this apparent mismatch with observations, the predicted column density distributions exhibit substantial scatter that brackets the observed distributions in both \HI\ and \MgII. Using an average of the simulated column density distribution therefore yields excellent agreement with observations, while the median column density lies well below them. As a result, drawing conclusions about the level of agreement between simulations and observations can be challenging when relying solely on column densities.
    
            \begin{figure}
                \includegraphics[width=\columnwidth]{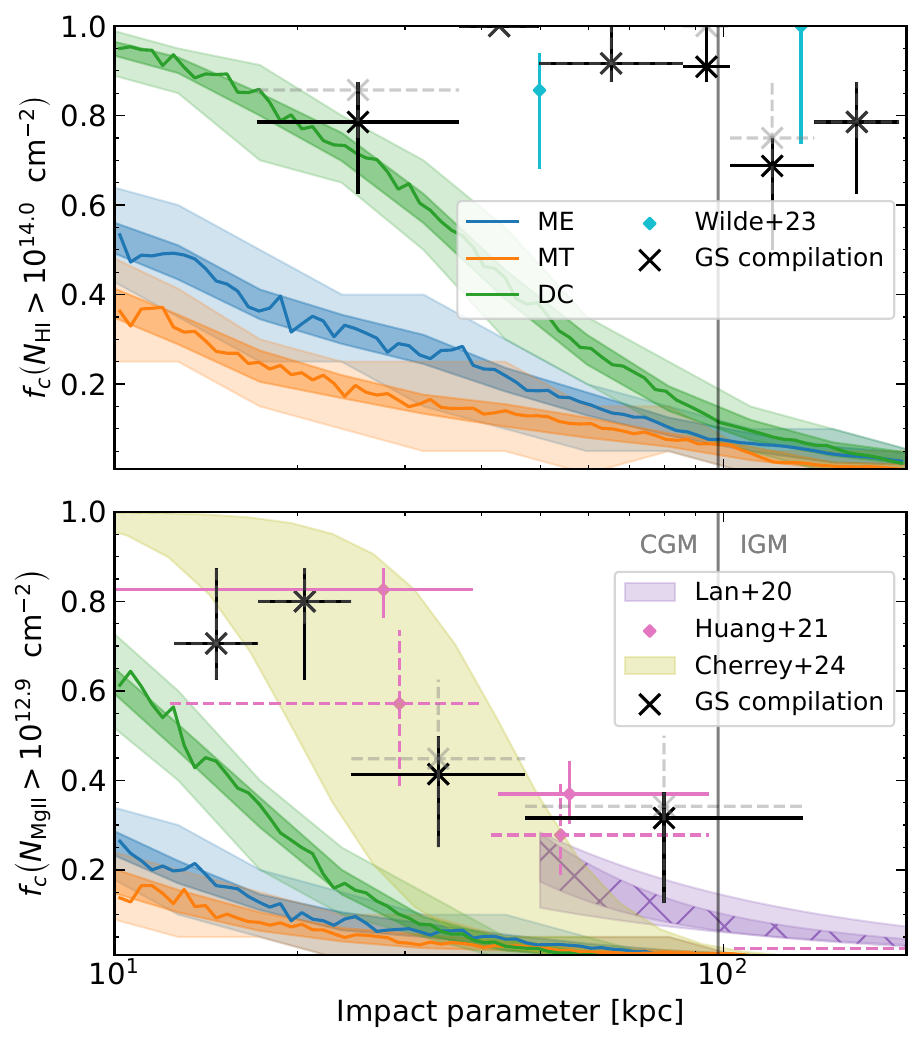}
            \caption{Covering fraction as a function of impact parameter for \HI\ (top) and \MgII\ (bottom) stacked over $1\rm\,Gyr$, from $z=1.3$ to $z=1$. The solid lines correspond to the binned fraction of column densities above the chosen threshold, with the same colour code as used in previous figures. We plot uncertainties estimated with bootstrapping on a subset of 240 (resp. 2400) column densities as a light-colored (resp. dark-colored) shaded area. We compare our simulated covering fractions to measurements from \citet{Wilde2023} for \HI, and to measurements from \citet{Lan2020}, \citet{Huang2021}, and \citet{Cherrey2025} for \MgII, as indicated in the legend. Star-forming and quiescent galaxies are respectively shown by the solid and dashed error bars in \citet{Huang2021}. In \citet{Lan2020}, passive galaxies are shown by the purple shaded area, while the hashed purple area shows star-forming galaxies. Finally, our compilation of galaxy-selected observations is shown with black crosses and error bars. The grey crosses with error bars are used in Appendix \ref{sec:Appendix_GS}. The vertical black line corresponds to the mean $R_{200}$ over our redshift range.}
            \label{fig:frac_cold}
            \end{figure}
    
            To overcome this challenge, and to account for the fact that many observational points are upper limits, we compute the \textit{covering fraction}, i.e. the fraction of column density measurements above a given threshold. If the surveys are complete above this limit, the covering fraction represents the fraction of simulated column densities exceeding the chosen detection threshold. This threshold must be high enough to remain above the observational detection limit, but not so high that the resulting statistics become too low. We discuss this limitation further in Section~\ref{subsec:limits_obs}.

            To estimate uncertainties in the measurement of covering fractions, we use bootstrapping to create subsets matched in size to observational samples. We also create another subset containing ten times more column densities. In practice, we randomly draw subsamples of 20 (resp. 200) measurements in each of 12 logarithmically spaced bins over $[8,240]\rm\,kpc$, and compute the covering fractions for the resampled dataset. This corresponds to datasets of 240 (resp. 2400) column densities, comparable to existing observational samples, such as \citet{Cherrey2025} and \citet{Huang2021}, which contain 127 and 211 measurements, respectively. We repeat this process 100 times, and define uncertainties using the 15.9th and 84.1th percentiles of the resulting distribution of covering fractions. Note that the number of bins used to compute the uncertainties of the covering fraction differs from that used for the median covering fractions. As a result, the median does not necessarily coincide with the uncertainties.
            Fig.~\ref{fig:frac_cold} shows the covering fraction of \HI\ and \MgII\ as a function of impact parameter with errors estimated by bootstrapping. We compare our simulated covering fractions to \citet{Wilde2023} for \HI, to \citet{Lan2020}, \citet{Huang2021}, and \citet{Cherrey2025} for \MgII, and to a compilation of our galaxy-selected surveys (hereafter \textit{GS compilation}) for both. We detail the GS compilation and our threshold selection in Appendix \ref{sec:Appendix_GS}. Briefly, for the simulations in \HI, we use a column density threshold of $N_\HI = 10^{14.1}\rm\,cm^{-2}$, close to that of \citet{Wilde2023}. For \MgII, we use an equivalent width threshold of $W_\MgII = 0.3\,\text{\AA}$ for the simulations, the GS compilation, \citet{Huang2021}, and \citet{Cherrey2025}, and we use $W_\MgII = 0.4\,\text{\AA}$ for \citet{Lan2020}. 

            In Fig.~\ref{fig:frac_cold}, we find that with a threshold of $N_\HI = 10^{14.1}\rm\,cm^{-2}$, all observed covering fractions in \HI\ are higher than $0.6$, significantly above the results from our simulations. Furthermore, within $R_{200}$, simulated covering fractions decrease steeply with increasing impact parameter, while observed covering fractions do not exhibit the decreasing trend that we would expect. This shows that the covering fraction for weak \HI\ absorbers extends much further than our simulations predict and that our \HI\ column densities are severely underestimated in all models.
            That being noted, observations seem to favour extremely strong feedback models such as that implemented in the \DC\ simulation. Although \citetalias{Rey2025} shows that \DC\ ejects roughly six times more of the metals it produces into the CGM than \KI\ and \KR, it expels $\approx64$\% of the metals produced out of the halo. This suggests that the model is overly efficient and that better agreement with observations may be obtained if more gas is retained within the virial radius rather than expelled to large distances.
            Although simulations and observations also showcase a discrepancy in \MgII, it is significantly smaller than for \HI\ owing to the much lower observed covering fractions. As for column densities, \DC\ exhibits higher covering fractions than the other two models in the inner CGM. Although the median column density in \KI\ exceeds that in \KR\ by $\sim$2 dex out to $\approx 15\rm\,kpc$ before dropping abruptly to comparable values, this trend is not mirrored in covering fractions. Covering fractions are primarly determined by the high-column-density tail of the distribution, as indicated by the 84.1 percentile: \KI\ maintains higher column densities than \KR\ out to $\approx 0.5\,R_{200}$, resulting in systematically higher covering fractions over the same range. Beyond $0.5\,R_{200}$, these differences disappear, as the column density threshold exceeds most values simulated in \KI\ and \KR. As a result, the covering fractions are similarly low in both models. We discuss such limitations in more detail in Section~\ref{sec:discussion}.

        \subsubsection{Warm gas: C\texorpdfstring{$\,\textsc{iv}$}{IV}}
            \begin{figure}
                    \includegraphics[width=\columnwidth]{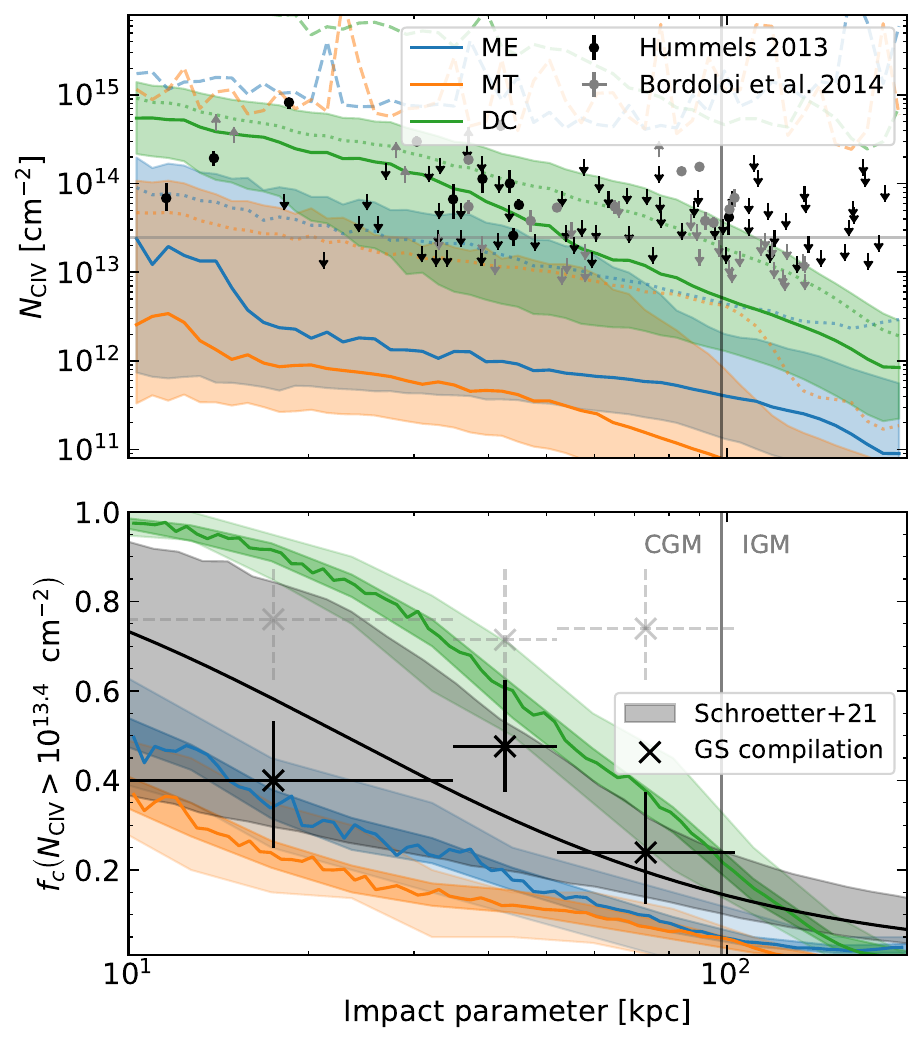}
            \caption{Column density (top) and covering fraction (bottom) of \CIV\ as a function of impact parameter stacked over $1\rm\,Gyr$, from $z=1.3$ to $z=1$. The same notation as Fig.~\ref{fig:rad_col_cold} (resp. Fig.~\ref{fig:frac_cold}) is used for the upper (resp. lower) panel. In the lower panel, we also show observations from \citet{Schroetter2021} (solid black line), with the shaded area corresponding to the $2\,\sigma$ confidence region, and the GS compilation.}
            \label{fig:rad_col_warm}
            \end{figure}
    
            We now look at the warm gas traced by \CIV\ and compare our column densities to observations from \citet{Hummels2013} and \citet{Bordoloi2014}.
            The upper panel of Fig.~\ref{fig:rad_col_warm} shows the median column density of \CIV\ as a function of impact parameter. \DC\ is 1-1.5 dex above \KI\ at all radii, which is in turn three to five times higher than \KR. The \KI\ and \KR\ profiles have similar shapes, unlike the \DC\ profile.
            Both the median and the mean column densities in \KI\ and \KR\ appear to be lower than the observed values, whereas \DC\ shows excellent agreement. However, once again, the simulated column densities cover a wide range of values and bracket observations for all simulations. Furthermore, most observations are upper limits, making comparisons with simulations uncertain.
    
            In the lower panel of Fig.~\ref{fig:rad_col_warm}, we show the covering fraction of \CIV\ as a function of impact parameter. We compare our results to \citet{Schroetter2021} and the GS compilation. 
            Although the GS compilation computed from galaxy-selected observations \citep{Hummels2013, Bordoloi2014} and the \citet{Schroetter2021} model are in good agreement, the discrepancy between DC and the GS compilation appears inconsistent with the seemingly excellent agreement shown in the upper panel. In \DC, the median column density crosses the column density threshold at $b\approx 50\rm\,kpc$. The covering fraction is thus above 0.5 at smaller impact parameters. Conversely, because we treat upper limits as non-detections, the scarcity of observed detections above the threshold results in a covering fraction below 0.5 at these impact parameters. We discuss this choice in Appendix~\ref{sec:Appendix_GS} and focus here on the comparison with the \citet{Schroetter2021} model.
            \DC\ exhibits column densities roughly twice as high as the other simulations and slightly above observations at all impact parameters. At distances larger than $\approx 50\rm\,kpc$, \DC\ shows a significantly steeper decrease in the covering fraction with distance than found in observations. In contrast, \KI\ and \KR\ are fairly close to the lower limit from observations. While \KI\ falls within the range of observations up to $\approx 50\rm\,kpc$, \KR\ shows significantly lower covering fractions.

        \subsubsection{Hot gas: O\texorpdfstring{$\,\textsc{vi}$}{VI}}
            \begin{figure}
                    \includegraphics[width=\columnwidth]{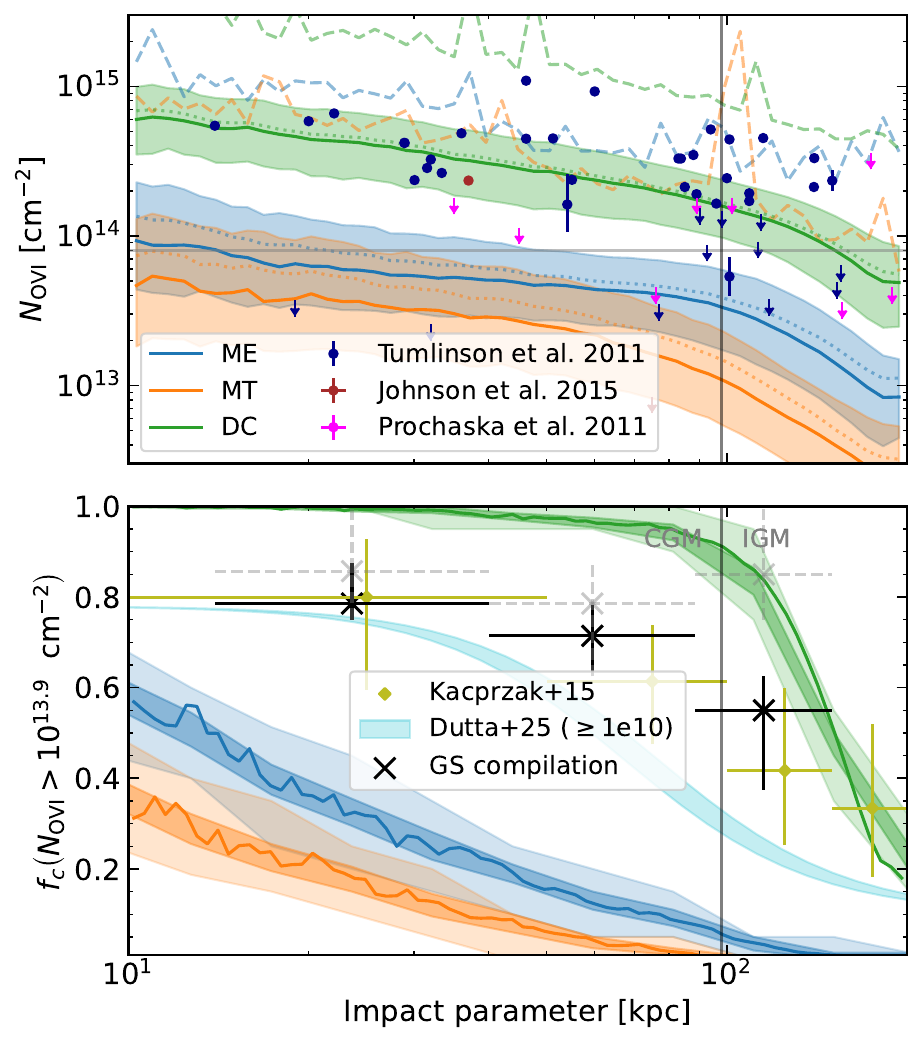}
            \caption{Column density (top) and covering fraction (bottom) of \OVI\ as a function of impact parameter stacked over $1\rm\,Gyr$, from $z=1.3$ to $z=1$. The same notation as Fig.~\ref{fig:rad_col_cold} (resp. Fig.~\ref{fig:frac_cold}) is used for the upper (resp. lower) panels. We also show observations from \citet{Tumlinson2011}, \citet{Johnson2015}, and \citet{Prochaska2011} in the upper panel, and observations from \citet{Kacprzak2015} and \citet{Dutta2025} in the lower panel.} 
            \label{fig:rad_col_hot}
            \end{figure}
    
            The last ion we examine is \OVI. The observational data we compare our results to are taken from \citet{Prochaska2011}, \citet{Tumlinson2011} and \citet{Johnson2015}. 
            The upper panel of Fig.~\ref{fig:rad_col_hot} shows the median column density of \OVI\ as a function of impact parameter. The three simulations show similar behaviour as for the other ions, with significantly higher column densities in \DC\ than in the other simulations, and somewhat higher in \KI\ than in \KR. The column density profiles show clear `knees' around $R_{200}$.
            \KI\ and \KR\ yield median column densities approximately three and five times lower than observations, respectively. Unlike previous ions, where simulations broadly encompass the observed values despite low medians, the \OVI\ distributions are narrow and only marginally bracket the data, though they still reach most observed values. On the other hand, \DC\ appears to show good agreement with observations. One should, however, note that most detections of \OVI\ shown come from a single survey \citep{Tumlinson2011}, while other observations are mainly upper limits \citep{Prochaska2011}.
    
            We now turn to the covering fractions presented in the lower panel of Fig.~\ref{fig:rad_col_hot}, and our results with those of \citet{Kacprzak2015} and \citet{Dutta2025}. We also include our own observational compilation, which is in excellent agreement with existing constraints.
            Most column densities in \DC\ exceed the column density threshold adopted for our covering fraction calculations (grey line in the upper panel). This results in covering fractions at or near unity, which remain systematically higher than observations out to $R_{200}$. \DC\ exhibits a similar decline as seen in the observations. Conversely, column densities in \KI\ and \KR\ are mostly below the grey line shown in the top panel and thus exhibit much lower covering fractions, and strong disagreement with observations.

        \subsubsection{Intermediate summary}
            \citetalias{Rey2025} has shown that the delayed cooling feedback model produces much stronger feedback than the other models, ejecting a significant amount of gas and metals into the CGM and beyond, leading to a massive CGM rich in metals. This results directly in higher column densities than with \KI\ and \KR\ in \MgII, \CIV\ and \OVI. Thus, except for the cold phase of the gas, \DC\ overestimates covering fractions compared to observations.
            The simulations using the \KI\ and \KR\ models exhibit distinct characteristics in their CGM: \KI\ features a gas-rich environment with low metallicity, and \KR\ shows a smaller quantity of high-metallicity gas. Despite these disparities, both models demonstrate an overall similar quantity of metals, although their distribution differs. \KI\ column densities and covering fractions are overall higher than \KR, and although both simulations exhibit an overall insufficient amount of gas in the CGM (i.e. both hot and cold), they are close to agreement with observed \CIV\ covering fractions.
            We highlight that there is a large scatter in the simulated column densities, which is stronger for low-energy ions. Crucially, reproducing the observed column densities, whether through the median, mean, or bracketing the data, does not guarantee a match in covering fractions. Simulations with a broad column-density distribution in which only a small fraction of sightlines exceed the column density threshold naturally produce low covering fractions compared to observations (e.g., \KI\ and \KR\ for \MgII). Conversely, a narrow distribution that appears consistent with observed column densities may place nearly all sightlines above the threshold, leading to excessively high covering fractions (e.g., \DC\ for \OVI). Given current observational limitations, comparing simulated and observed column densities serves only as a first-order diagnostic. Covering fractions offer a more stringent and discriminating constraint, as they probe the full statistical distribution of sightlines rather than relying on the specific values of individual detections.

 \section{Discussion} \label{sec:discussion}
    In the previous section, we have shown key differences in the three simulations and the properties of their CGM. We now discuss the caveats and limitations of our comparisons, and which improvements can be made to reach better agreement between observations and the simulations presented here.

    \subsection{Uncertainties in the comparison with observations} \label{subsec:limits_obs}
        \subsubsection*{Low statistics of the column densities}
            Galaxy-selected surveys are scarce, and the statistics we have from observations are relatively low. \MgII\ and \CIV\ column densities are only probed by two surveys, and for each ion, one of the two surveys relies on strong priors \citep{Chen2001, Hummels2013}. The surveys for both of these ions are clearly limited by sensitivity, as we find a plateau of upper limits for both of them. We find a similar issue with \OVI, which is probed by three surveys, and yet, all detections but one come from the same survey \citep{Tumlinson2011} and cover a very narrow range of column densities. Even in covering fractions, \HI\ and \CIV\ only have one observational paper to be compared against. \citet{Nielsen2013} also provides \CIV\ covering fractions, but we chose not to include them as multi-object spectrographs cannot resolve spectra of sources denser than a few per arcmin. In contrast, \citet{Schroetter2021} relies on Multi-Unit Spectroscopic Explorer (MUSE) data, which provides the spatial resolution required to distinguish closely-spaced sources. More surveys with high completeness are thus a key component to a better understanding of the CGM.

        \subsubsection*{Mass and redshift dependence}
            Studies suggest that there is a strong positive correlation between the absorption strength/size of the cold CGM and the galaxy or halo mass \citep{Bordoloi2018, Wilde2021}. The galaxy simulated in this study has a stellar mass of $M_\star\approx 10^{10.2}\rm\,M_\odot$. The galaxies for which we have observational constraints cover a very broad range of masses between $10^{9.5}\,\mathrm{M_\odot} < M_\star < 10^{11.5}\,\mathrm{M_\odot}$. If the correlation between galaxy mass and absorption strength is indeed real, the discrepancy we find between simulations and observations may be partially driven by stellar mass.

            In \MgII, all galaxy-selected column density measurements used in this study correspond to redshifts $z<0.4$ (see Appendix~\ref{sec:Appendix_GS}). As a result, any redshift dependence in column density may influence our conclusions. Indeed, \citet{Lundgren2021} reports a significant evolution in the physical extent of \MgII. However, this conclusion is based on comparisons with \citet{Nielsen2013} and \citet{Dutta2020}, which were obtained through a heterogeneous selection. Furthermore, while \citet{Dutta2020} finds no dependence on the equivalent width or covering fractions over $z\sim0.3-1.5$, \citet{Nielsen2013} relies on a multi-unit spectrograph, which is less robust than IFU-based observations. Thus, the apparent evolution reported by \citet{Lundgren2021} may be considered tentative.
            Using the DESI Legacy survey, \citet{Lan2020} found no redshift evolution in the covering fractions of weak absorbers, a result confirmed by \citet{Chen2025} using the DESI EDR. With the DESI DR1, \citet{Wu2025} found little evolution in redshift, possibly driven by their evolution in star formation rate or stellar mass. However, DESI select galaxies with a density of $400\rm\,deg^{-2}$ \citep{Raichoor2023}, while MUSE reaches $\sim50\rm\,arcmin^{-2}$. DESI is restricted to the most massive, luminous systems, while MUSE captures a much broader and more complete galaxy population. Not being complete in mass over all redshift ranges could explain the absence of evolution with redshift in the DESI results, as the evolution may be mass-dependent.
            With MUSE, the MUSE Analysis of Gas around Galaxies (MAGG) survey \citep{Dutta2021} and the DR2 of the MUSE GAs FLOw and Wind (MEGAFLOW) survey \citep{Cherrey2025} both find a tentative evolution of the \MgII\ covering fractions for weak absorbers with redshift. However, column densities in \citet{Dutta2021} show no strong trend with redshift, and the covering fraction evolution in $(1+z)^{0.77}$ found by \citet{Cherrey2025} has uncertainties compatible with no evolution. Furthermore, these two surveys also suffer from caveats. While the MAGG survey is deficient in strong absorbers, MEGAFLOW DR2 appears to be somehow lacking in weak absorbers \citep[see Fig.~4 in][]{Bouche2025}. This could enhance the apparent redshift dependence, as \citet{Lan2020} found that strong absorbers exhibit a more pronounced evolution with redshift compared to weak absorbers.
            Overall, the redshift dependence does not appear to be significant. Even if the column densities evolve with redshift as $(1+z)$, this would not impact our conclusions, as the discrepancy between observations and simulations in \MgII\ is substantially larger.

        \subsubsection*{Covering fractions and $W_\mathrm{min}$}
            \begin{figure}
                \includegraphics[width=\columnwidth]{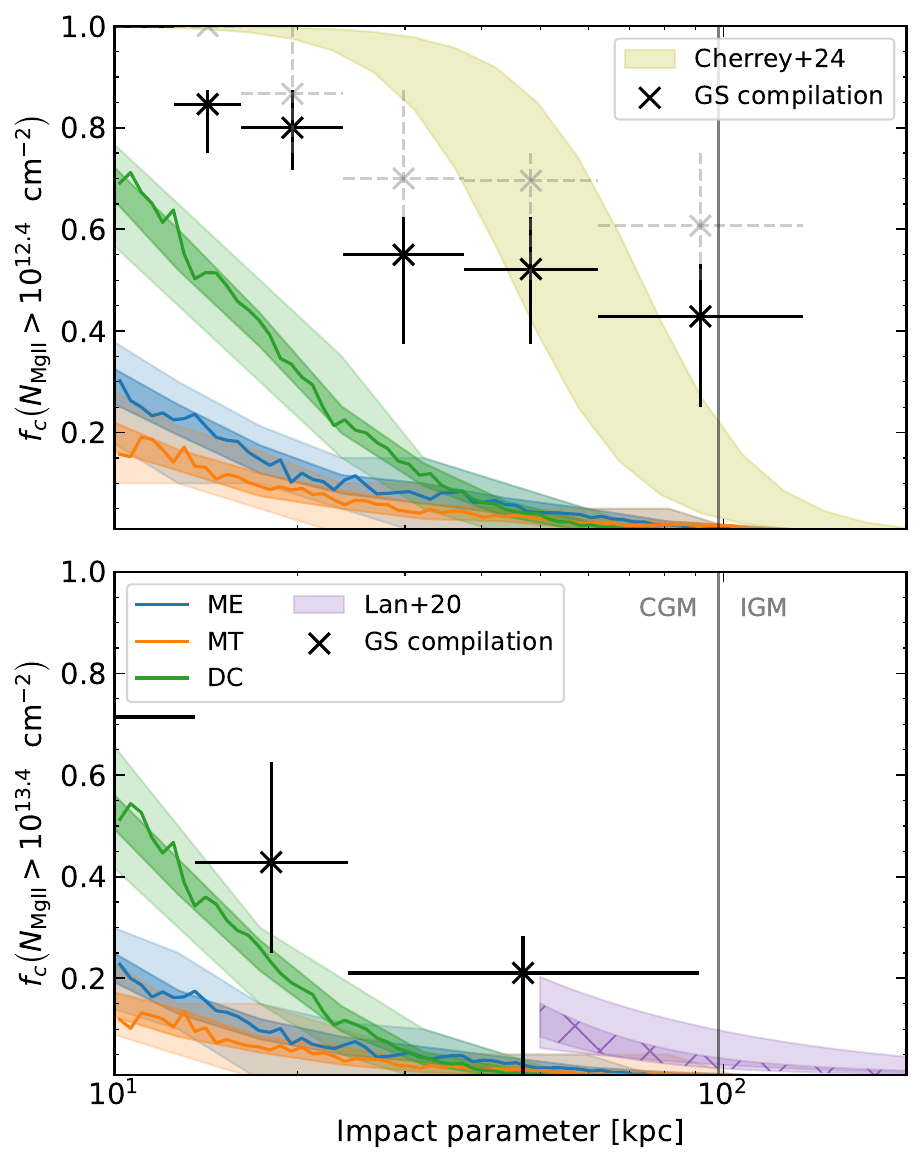}
            \caption{Covering fraction as a function of impact parameter for \MgII\ stacked over $1\rm\,Gyr$, from $z=1.3$ to $z=1$. The column density thresholds correspond to $W_\MgII = 0.1\,\text{\AA}$ (top), and $W_\MgII = 1.0\,\text{\AA}$ (bottom). An equivalent width of 0.1Å corresponds to a column density of $2.39\times10^{12}\rm\,cm^{-2}$ in \MgII\ which is slightly lower than the minimum value in the GS compilation, but this difference has no impact on our findings. 
            The simulated data are shown as coloured solid lines with the same colour code as used in the previous plots. We compare the \MgII\ covering fractions to the GS compilation (black crosses) and \citet{Cherrey2025} (yellow shaded area) in the top panel. In the bottom panel, we compare our results to the GS compilation (solid black and dashed grey crosses) and \citet{Lan2020} for passive (purple shaded area) and star-forming (hashed purple area) galaxies.}
            \label{fig:frac_cold2}
            \end{figure}
            To address the limitations of column density measurements (limited in sensitivity, rare galaxy-selected surveys), one can turn to the covering fractions \textit{given that the selected threshold is high enough}. 
            We have shown in the bottom panel of Fig.~\ref{fig:frac_cold} that the covering fractions of \MgII\ in our simulations are lower than those from observations, using a column density threshold corresponding to an equivalent width of $W_\MgII = 0.3\,\text{\AA}$. In Fig.~\ref{fig:frac_cold2}, we show the same quantity, except that we change the threshold to $W_\MgII = 0.1\,\text{\AA}$ (top panel) and $W_\MgII = 1.0\,\text{\AA}$ (bottom panel). We also update our comparison points to studies matching these thresholds. We compare our result to the GS compilation, \citet{Cherrey2025} for $W_\MgII = 0.1\,\text{\AA}$, and to the GS compilation, and \citet{Lan2020} for $W_\MgII = 1\,\text{\AA}$, as in the previous section. Comparing the top and bottom panels, we observe a significant decrease in covering fractions across all radii and for all observations when the threshold is increased. 
            However, the \MgII\ covering fractions derived from simulations exhibit a much slower decrease than observations, bringing the simulation points in or close to agreement with observations when using the higher column density threshold. The slower decrease observed in simulations can be attributed to both significantly higher statistics than observations and the lower column density threshold approaching the sensitivity limit of observations. Notably, all simulations are much closer to observations with a high column density threshold, whereas these covering fractions are significantly underpredicted with a lower column density threshold, as seen in the top panel. This discrepancy arises because fewer strong absorbers are expected compared to weak absorbers, resulting in lower covering fractions for strong absorbers, which align more closely with simulation results. This comparison highlights that at least for strong absorbers, simulations better reproduce observations, possibly because they are not associated with feedback but rather tidal features. We argue that comparisons focusing on weak absorbers are essential for placing constraints on feedback models.

            \begin{figure}
                \centering 
                \includegraphics[width=\columnwidth]{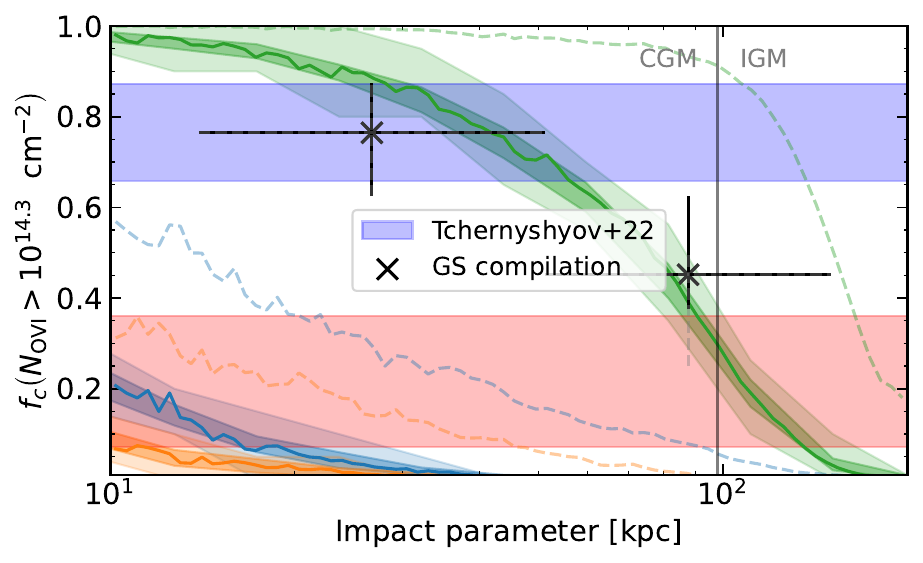}
                \caption{Covering fraction of \OVI\ as a function of impact parameter stacked over $1\rm\,Gyr$, from $z=1.3$ to $z=1$ with a column density threshold of $N_\OVI = 10^{14.3}\rm\,cm^{-2}$. The notation is the same as in Fig.~\ref{fig:frac_cold}. We also show covering fractions from \citet{Tchernyshyov2022} for active (blue) and passive (red) galaxies. We additionally show in dashed lines the covering fractions and GS compilation points obtained with a density threshold of $W_\OVI \approx 0.1\,\text{\AA}$, as done in Fig.~\ref{fig:rad_col_hot}.}
            \label{fig:rad_col_hot2}
            \end{figure}

            \vspace{0.3 cm}
            
            The expected range of column densities probed by a given ion should also play a significant role in determining the threshold used to calculate covering fractions. Unlike \MgII\ or \HI, the spread in column densities in \OVI\ is very narrow for all models, covering less than 1 dex. This limited spread, combined with little correlation with impact parameter, results in a very high sensitivity of the covering fraction to the threshold. A small change in the threshold can result in most points being removed, drastically affecting the covering fractions. 

            In Fig.~\ref{fig:rad_col_hot2}, we show the covering fraction of \OVI\ using a column density of $N_\OVI = 10^{14.3}\rm\,cm^{-2}$ ($W_\OVI \approx 0.25\,\text{\AA}$). This threshold corresponds to observations from \citet{Tchernyshyov2022} and is $\approx 1.6$ times higher than in Fig.~\ref{fig:rad_col_hot} in column density ($W_\OVI \approx 0.1\,\text{\AA}$). Using this higher threshold has little effect on the GS compilation, except that two points are merged into one as per our lower limit of 10 points per bin. However, the threshold selection affects the simulations considerably, bringing their covering fractions down to much lower values. \DC\ now finds very good agreement with observations in the outer CGM, while \KI\ and \KR\ go to even worse agreement with observations than before. Fig.~10 from \citet{DeFelippis2024} also shows quantitatively how the impact parameter at which the covering fraction reaches 50\% varies as a function of the column density threshold for simulations sharing the same initial conditions. For high-energy ions such as \CIV\ or \OVI, there is a strong dependence on the threshold used, as the range of column densities in the CGM is narrow. For low-energy ions such as \MgII, this dependence is small due to a large scatter in simulated column densities.

    \subsection{Why are CGM observables challenging to reproduce?}
        
        \subsubsection*{The integration length}
            \begin{figure}
                \centering
                \includegraphics[width=\columnwidth]{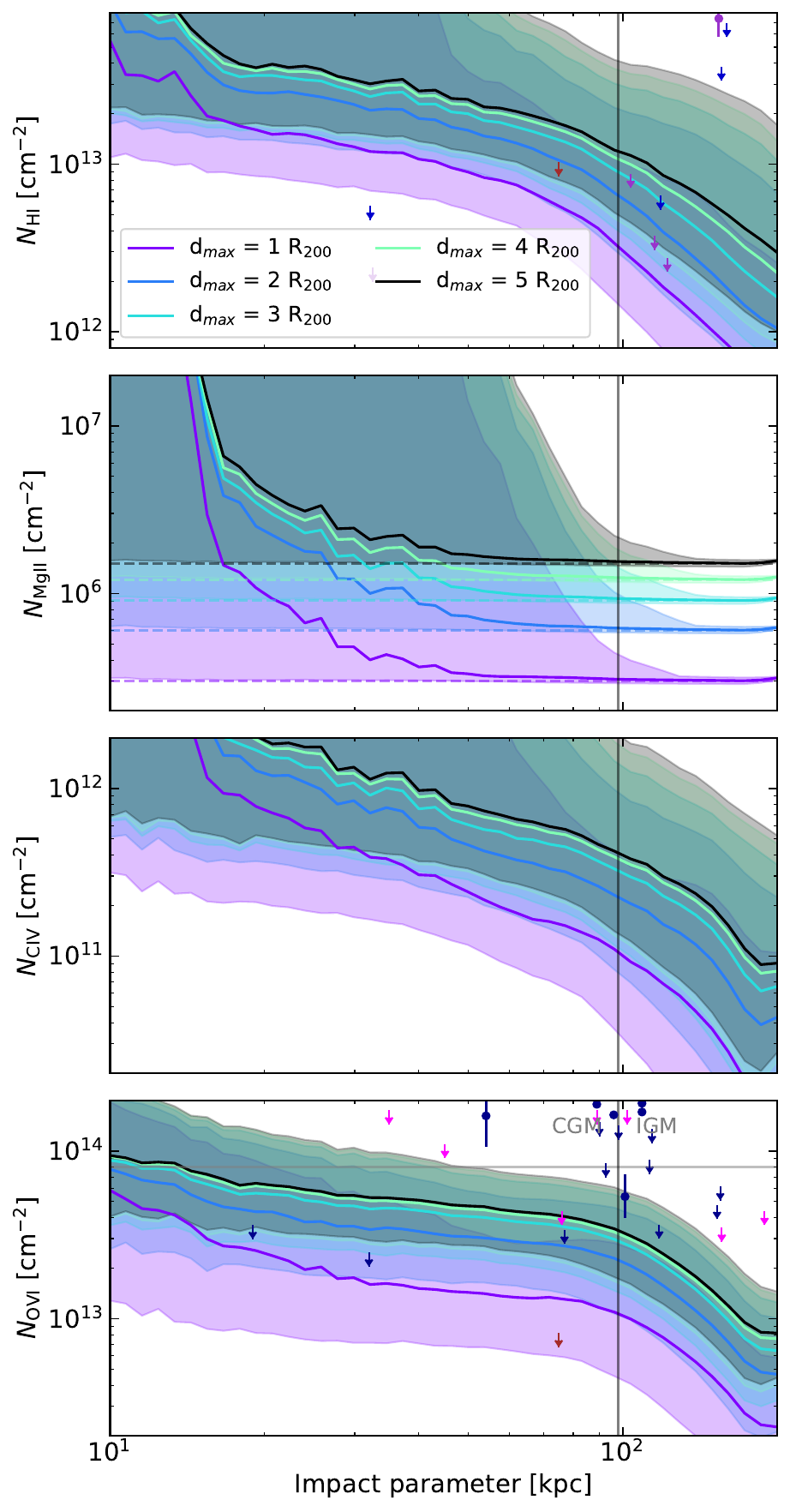}
                \caption{
                Column density in \KI\ as a function of impact parameter for \HI, \MgII, \CIV, and \OVI\ (from top to bottom), stacked over $1\rm\,Gyr$, from $z=1.3$ to $z=1$. Except for $5\,R_{200}$, which is shown in black, the different colours correspond to different integration lengths, from $R_{200}$ (purple) to $6\,R_{200}$ (turquoise). The solid lines and shaded areas follow the same definition as the previous plots, and observations are the same as in Fig.~\ref{fig:rad_col_cold}, ~\ref{fig:frac_cold}, ~\ref{fig:rad_col_warm}, and ~\ref{fig:rad_col_hot}. The dashed lines represent the numerical floor discussed in the text.}
                \label{fig:artifdmax}
            \end{figure}

            To compute column densities, we integrate gas densities over a $5\,R_{200}$ path length. This depth is comparable to observational integration limits and is centred on a plane intersecting the galactic centre perpendicularly to the line-of-sight. Within this integration length, it is typically assumed that the CGM dominates the signal and that contributions from the surrounding IGM are negligible. However, as we demonstrate in Fig.~\ref{fig:artifdmax}, the contribution from the diffuse IGM is not negligible. We use this figure to illustrate the impact of the IGM and quantify the sensitivity of our results to the choice of the integration length. 

            In Fig.~\ref{fig:artifdmax}, we show \HI, \MgII, \CIV, and \OVI\ column densities for integration lengths in the \KI\ simulation.  ranging from $R_{200}$ to $5\,R_{200}$ (our default) in the \KI\ simulation. 
            We show here that all four ions depend on the adopted integration length: column densities measured by integrating over $5\,R_{200}$ along the sightlines are a few times higher than those measured when integrating only over $R_{200}$. In all cases, however, the column densities converge as the integration length increases and remain well below the observed values. Therefore, our choice of integration length has no significant impact on our conclusions. 

            While the integration length does not significantly affect our conclusions, the scale at which the column densities converge provides insight into the physical location of the absorbing gas along the sightline. 
            For a sightline at an impact parameter of $R_{200}$, an integration path length of $L\,R_{200}$ samples absorption from gas at physical distances up to $\sqrt{\left({L}/{2}\right)^2+1}\, R_{200}$ from the galactic centre\footnote{An integration length of $L\,R_{200}$ means that the gas contributing to the absorption spans $\pm\,\frac{L}{2}R_{200}$ perpendicularly to the axis going from the absorption to the centre of the galaxy.}. 
            Within the CGM, gas along the sightline contributes appreciably to \HI, \CIV, and \OVI\ absorption out to $3-4,R_{200}$, indicating that material up to $\sim2\,R_{200}$ from the galactic centre contributes to the measured absorption. In turn, this indicates that a significant fraction of the absorbing gas originates from the outer CGM or the close IGM.
            The spatial distribution of \CIV\ and \OVI\ absorbers is consistent with Fig.~\ref{fig:mapsions}, which shows that these ions reside in a hotter ambient CGM phase than low ionisation species, with \CIV\ preferentially envelopping filamentary structures, and \OVI\ being more homogeneously distributed. The additional contribution from gas out to $2\,R_{200}$ is in excellent agreement with the findings of \citet{Qu2023}, who showed that intermediate- and high-ionisation species are more spatially extended than low-ionisation species, remaining detectable out to $2\,R_\mathrm{vir}$. 
            Although our results for \HI\ may appear to contradict their finding that low-ionisation species are confined to the inner halo ($\lesssim 0.8\rm\,R_{vir}$), this apparent discrepancy primarily reflects differences in the column density regimes probed. The \HI\ absorbers analysed here predominantly have column densities below $\sim10^{14}\rm\,cm^{-2}$, whereas the sample of \citet{Qu2023} consists exclusively of strong absorbers with column densities exceeding $\sim10^{16}\rm\,cm^{-2}$.
            At impact parameters larger than $R_{200}$, \HI\ column densities begin to diverge, while those of metal ions remain converged. This reflects the fact that \HI\ extends far beyond the halo, tracing distant structures such as cosmic filaments, whereas \CIV\ and \OVI\ require metals expelled by the galaxy. As the metal content declines with distance, the additional contribution from larger integration length becomes negligible, producing converged curves. The divergence in \HI, however, is small and does not affect our conclusions, as the column densities remain well below observational values.

            The behaviour of \MgII\ column densities as a function of the integration length contrasts with that of other ions, increasing appreciably up to $5\,R_{200}$ at impact parameters larger than $\approx 50\rm\,kpc$. Close to the galaxy, column densities are dominated by dense gas in the inner halo, which far exceeds contributions from the outer halo, so extending the integration length has no effect. The direct contribution from the galaxy sharply decreases at larger impact parameters, and the effect of longer integration lengths becomes more noticeable, with curves slowly diverging from $\approx 20\rm\,kpc$ to $\approx 50\rm\,kpc$. At these distances, \MgII\ originates primarily from satellites and their tidal tails, which act as reservoirs of enriched, cold gas. Further than $\approx 50\rm\,kpc$, the \MgII\ column densities plateau. This behaviour reflects a numerical limitation rather than a physical trend. In our post-processing, ion densities are floored at ${\sim10^{-18}\rm\,cm^{-3}}$. When integrated over $3\,R_{200} \approx 300\rm\,kpc$, this floor corresponds to column densities of $\sim10^{6}\rm\,cm^{-2}$, matching the numerical floor visible in Fig.~\ref{fig:artifdmax}. As a result, increasing the integration length simply scales these column densities by a constant factor. 
            Therefore, our result indicates that at least half of the sightlines contain negligible \MgII, producing a strongly bimodal distribution. This pronounced dichotomy, where most sightlines are essentially devoid of \MgII\ while a small fraction exhibit much higher column densities, reveals the highly patchy nature of \MgII. This patchiness explains the large discrepancy between observations and the bulk of simulated sightlines.

            Overall, all ions reasonably converge within an integration length of $5\,R_{200}$, and varying the integration length does not impact our results. \HI\ is broadly distributed through the CGM, while \CIV\ and \OVI\ mainly originate from the outer CGM and inner IGM. In contrast, most sightlines at large distances contain little or no \MgII.


        \subsubsection*{Are there enough metals in the CGM?}
            \begin{figure}
                \includegraphics[width=\columnwidth]{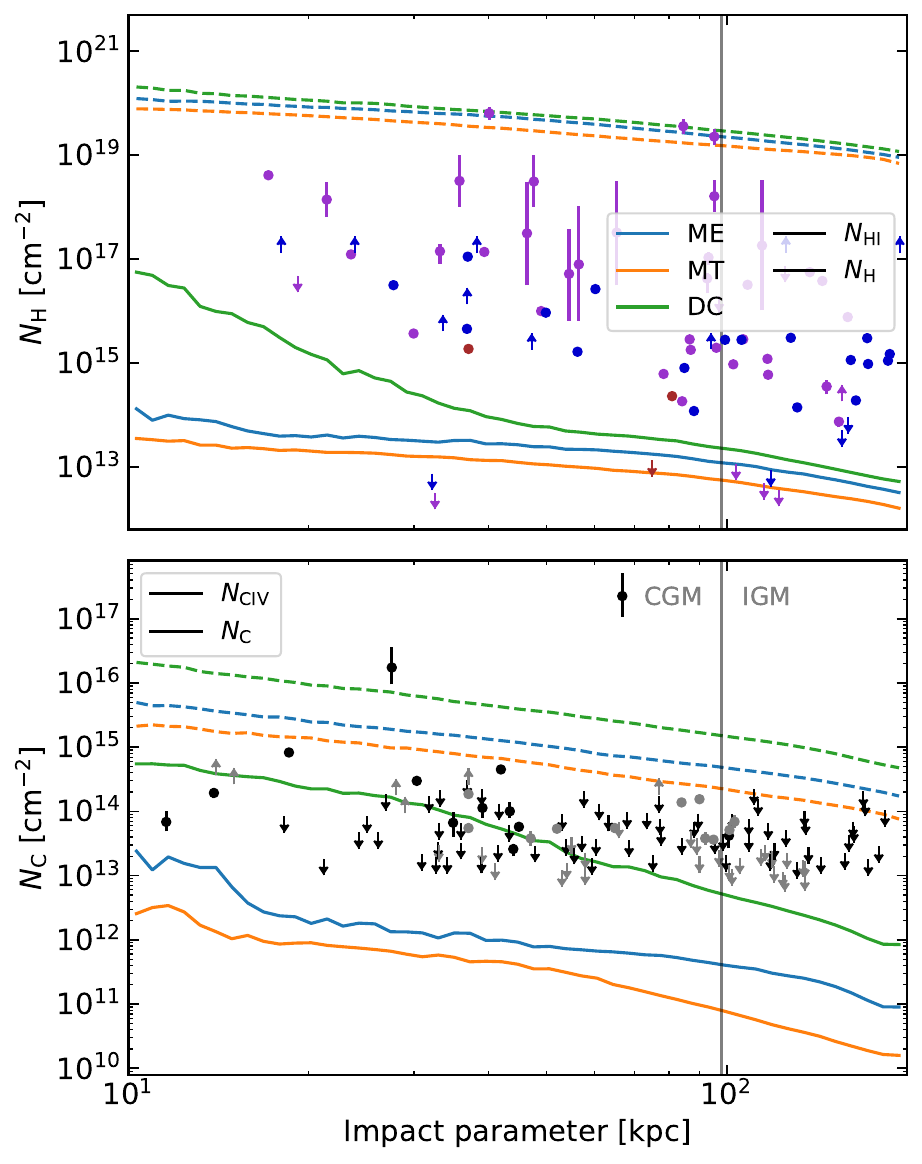}
                \caption{Column density of \HI\ (top) and \CIV\ (bottom) as a function of impact parameter stacked over $1\rm\,Gyr$, from $z=1.3$ to $z=1$. We also show in dotted lines the column densities of H and C, which correspond to the highest column density attainable for each ion, given the current amount of metals in the gas. Otherwise, the same notation and colour code are used as in the previous figures.}
                \label{fig:col_max}
            \end{figure}
    
            \vspace{0.3 cm}
            Based on Fig.~11 from \citetalias{Rey2025}, we argued that the higher metal content in the CGM of \DC\ was the primary driver for its elevated \CIV\ and \OVI\ column densities relative to \KI\ and \KR. Indeed, \citetalias{Rey2025} showed that the CGM of \DC\ contains approximately $2.5$ times more metals than that of the other two simulations. A comparison of the corresponding \CIV\ and \OVI\ covering fractions indicates that simply increasing the efficiency of metal transport into the CGM is sufficient to reproduce the observed level for these ions. The situation is markedly different for \MgII. Observational constraints in Fig.~\ref{fig:frac_cold2} show that the \MgII\ covering fraction reaches 0.5 at an impact parameter of $\approx30\rm\,kpc$. Reproducing this value would require the simulated median column density to reach the covering fraction threshold of $N_\MgII = 10^{12.9}\rm\,cm^{-2}$ at the same distance. Instead, the median column density in \KI\ and \KR\ (resp. \DC) falls short of this threshold by roughly 6 dex (resp. 3 dex). Bridging this gap through an increase in CGM metallicity alone would require an implausibly large enhancement, far exceeding any reasonable uncertainty. This demonstrates that, for \MgII, the total metal mass in the CGM is unlikely to be the primary limiting factor in reproducing the observations. Nevertheless, it remains unclear whether the existing CGM metal budget could yield the observed \MgII\ properties under different physical conditions. We explore this possibility in the following section.

            In Fig.~\ref{fig:col_max}, we show the column densities of H and \HI\ in the upper panel, and C and \CIV\ in the lower panel (the results are similar for \MgII\ and \OVI). For all species considered, the column density obtained with the atom instead of a specific ion is far above most observed values, and the covering fractions (not shown) are all at unity up to the highest impact parameter probed. The column density of each element can be interpreted as the integrated density of the element including all its ionisation states, but also as an upper limit for any ionic column density (\HI\ and \CIV\ in our example), corresponding to $f_{\rm ion}=1$. This simple shift in perspective shows that even though increasing the amount of gas or metals in the CGM would improve the agreement between observations and simulations, \textit{there is in principle sufficient gas and metals in the CGM to match observed column densities}.
    
            Independent of metallicity, the $\sim6$ orders of magnitude difference between $N_\mathrm{H}$ and $N_\HI$ indicates that the CGM of simulated galaxies is indeed highly ionised. To match the observations, we can see from Fig.~\ref{fig:col_max} that the mean neutral fraction must increase by a factor $\sim10-100$, shifting from $f_\mathrm{ion}\sim10^{-6}$ to $f_\mathrm{ion}\sim10^{-4}$ (we approximate the ionisation fraction as the ratio of the two curves). Such a shift could be achieved, for example, by changing the amplitude of the UV background in our simulation. Similarly, the fraction of \MgII\ and \CIV\ should ideally reach $f_\mathrm{ion}\sim10^{-2}$, whereas for \OVI, the optimal ionisation fraction required to match observations is found to be less than $f_\mathrm{ion}\sim10^{-1}$. For both ions, while a modest increase in the CGM metal mass is sufficient to reach agreement with observations, small adjustments to their ionisation fractions could yield a similar improvement.

            In conclusion, we find that while subgrid models ejecting slightly more metals into the CGM (or fewer, in the case of \DC) would readily produce a better match with observations for \CIV\ and \OVI, such adjustments would be largely insufficient for \MgII. For both \HI\ and \MgII, the primary discrepancy in our simulations appears to stem from their ionisation fractions rather than the total metal content.


    \subsection{Uncertainties in the modelling}  \label{subsec:uncertainties_sims}
        We have shown that the comparison between simulations and observations is subject to several caveats, including limitations in sensitivity and statistics, as well as the inherent complexity in choosing appropriate thresholds for covering fractions. However, even taking these effects into account, the simulations still exhibit an amount of \HI, \MgII, and \OVI\ globally lower than observations, especially for \KI\ and \KR. We have seen that possible shortcomings of the simulations are likely to come from the galaxies being inefficient in ejecting enough metals or in the ionisation state being incorrectly modelled. While a change in ionisation fraction seems necessary for \MgII, increasing the metallicity by small amounts can bring the simulated \CIV\ and \OVI\ in good agreement with observations, while it seems largely insufficient for \MgII. We now examine what might cause these discrepancies in simulations and how to alleviate them.

        \subsubsection*{Non-equilibrium thermochemistry}
            Several sources of uncertainty lie in our post-processing of the ion abundances. 
            The first lies in the radiation field. Ideally, the radiation field in our simulation would cover all energies with as little discretisation as possible. However, as this is unfeasible, we chose to reduce it to the three most important bins for \HI, \HeI, and \HeII. We omitted non-ionising energies from our radiation field as they primarily affect \ion{Mg}{i}. This simplification is justified given that \ion{Mg}{i} is expected to be negligible in the highly ionised environment of the CGM.
            A larger uncertainty lies in the UVB modelling. \citet{Mallik2023} found that changing the UVB can impact CGM observables as much as a change in the feedback recipe. Similarly, \citet{Taira2025b} found that the bulk of their column densities in \HI\ (resp. \CIV\ and \OVI) could vary by up to 4 dex (resp. 0.25 and 1 dex) depending on the chosen UVB. Nonetheless, these models typically omit local stellar radiation, which has been shown to dominate the CGM within $\approx 50\rm\,kpc$ for galaxies similar to ours \citep{Werk2014, Chen2017b}. \citet{Zhu2024} even found that stellar radiation dominates the CGM at all redshifts in their simulations. In our simulations, we find that local radiation dominates over the UVB in the inner CGM in the energy bin responsible for \CIV\ and \OVI\ photoionisation. Because these two ions primarily originate in the outer CGM, the findings of \citet{Mallik2023} and \citet{Taira2025b} should not be affected by the presence of local stellar radiation. Conversely, for the bins responsible for \HI\ and \MgII\ photoionisation, local stellar radiation dominates the flux throughout the entire halo. This prevalence makes it challenging to quantify how sensitive these lower-ionisation species are to the specific choice of UVB.

            Another source of uncertainty lies in the post-processing of the ions. At present, one of the most efficient ways to compute ion fractions from density, metallicity, temperature and ionising radiation is the method from \citet{Mauerhofer2021b}. Extrapolating CLOUDY tables is the standard approach for simulations without radiative transfer. 
            In our case, accounting for a spatially varying radiation field would increase the computational cost by a factor of $\sim10^4$ compared to our current solution \citep{Mauerhofer2021}. The physically preferable approach would be to evolve a full non-equilibrium thermochemical network on-the-fly, as done in \citet{Katz2024}, although this is still computationally demanding in both CPU time and memory footprint \citep{Katz2022c}. Recently, \citet{Kumar2025} demonstrated that assuming photoionisation equilibrium can underestimate or overestimate the fractions of high ions such as \CIV\ and \OVI\ by factors of a few in metal-enriched, rapidly cooling gas. However, in the low-density, high-temperature CGM regime where \OVI\ is predominantly collisionally ionised, the deviation from equilibrium is likely less dramatic.
            \citet{Cadiou2025} recently ran a simulation similar to ours with local stellar radiation coupled to a non-equilibrium chemical network. In addition, they track metal species and model non-ionising energy bins on-the-fly. While they found a modest effect when comparing their simulation to a post-processed version of their simulation assuming photoionisation equilibrium (i.e. without stellar radiation), this test does not isolate the impact of non-equilibrium thermochemistry and might be dominated by the removal of radiation. Indeed, the only other simulations we are aware of that explicitly study the effect of radiation on the CGM show that it can suppress \HI\ and \MgII\ column densities by 1 dex in the inner CGM \citep{Zhu2024}. As a result, it remains difficult to draw robust quantitative conclusions about the specific role of non-equilibrium thermochemistry.

            Although ion modelling uncertainties exist, their impact is likely subdominant compared to the uncertainties inherent in star formation and feedback prescriptions. These models not only regulate the mass and energy exchange with the CGM, but also reshape the local radiation field by altering the gas distribution around star-forming regions and thus the ionising photon escape fraction.

        \subsubsection*{Additional physics: AGNs}
            The ejection of metals in simulations is directly linked to the feedback processes. If the discrepancy between simulations and observations in \CIV\ and \OVI is due to a metal deficiency in the CGM, we need models that eject metals more efficiently from the galaxy.
            As seen with \DC, the amount of metals ejected plays a crucial role in reproducing observed \OVI\ column densities. Results from \citet{Tumlinson2011} and \citet{Tchernyshyov2022} suggest that \OVI\ column densities depend on the star formation activity of the host galaxy, with quiescent systems exhibiting significantly lower values than star-forming ones (see \citet{Tchernyshyov2022}, Fig.~\ref{fig:rad_col_hot2}). However, \citet{Nelson2018} argues that this correlation is a direct impact from black hole feedback, which can simultaneously increase \OVI\ column densities in the CGM and drive galaxy quenching \citep{Sharma2022, Bluck2023}. Indeed, it was found that in galaxies with stellar masses comparable to ours that including AGN could increase \OVI\ column densities by 1.5 dex \citep{Sanchez2019} and allow for a better match with observations. There are two main explanations for such an increase. Firstly, AGNs can greatly enrich the CGM by effectively expelling more metals out of the galaxy \citep{Eisenreich2017, Nelson2018, Sanchez2024}. Secondly, AGNs can also heat the CGM to temperatures corresponding to the ionisation fraction peak of \OVI\ ($10^{5.5-5.8}\rm\,K$) \citep{Suresh2017} and photoionise gas through, for example, their flickering \citep{Oppenheimer2013, Oppenheimer2018}, increasing \OVI\ column densities. Including AGNs in simulations thus seems essential to reproduce hot CGM observables.
            Crucially, such feedback is unlikely to worsen the comparison for \HI\ and \MgII, as these species trace a distinct, cooler gas phase. Recent studies indicate that cold clouds can survive when entrained in hot winds \citep{Gronke2022, Villares2024, Dutta2025b, Yao2025}. Consequently, cloud survival under more energetic feedback is likely a secondary concern for our current simulations compared to other bottlenecks, which we address in the following sections.

        \subsubsection*{Resolution}
            We have seen that both \MgII\ and \HI\ return similar results and fail to reproduce observational constraints. \HI\ is modelled on the fly with a dedicated ionisation bin, and we thus expect that our photoionisation modelling is reasonable. Furthermore, \HI\ does not depend on metallicity and a large amount of gas is available in all our simulations. A too-low ionisation fraction could thus be the main cause for these discrepancies.
            Finding the origin of too-low ionisation fractions is not trivial. Recent studies suggest that the cold CGM could be in the form of cold clouds with sizes of $\sim1-100\rm\,pc$ \citep{McCourt2018, Faucher-Giguere2023}, unresolved by current numerical simulations and thus spuriously mixed in the ambient hot medium. An increase in resolution could thus lead to better agreement with observations \citep{vandeVoort2019, Suresh2019, Peeples2019, Hummels2019}. However, the \ARCHITECT\ simulations already feature some of the highest resolutions currently available, with a median cell size of $0.3-2\rm\,kpc$ throughout the CGM (see \citetalias{Rey2025}, Fig.~1). Despite this, our resolution remains significantly coarser than the $1-100\rm\,pc$ scales expected for cold clouds in the CGM. More importantly, the effects of increased resolution on cold gas are still not converged or fully understood, and it is not clear that a higher resolution will solve the problem. For example, \citet{Rey2024b} have recently shown that resolving the cooling scale down to $18\rm\,pc$ in the CGM does not significantly change the covering factor of \HI, \MgII, or \CIV\ gas above column densities of interest. 

        \subsubsection*{Additional physics: cosmic rays}
            Recently, cosmic rays (CRs) have gained attention as another source of feedback \citep{Jubelgas2008, Pfrommer2017, Chan2019}. CRs can influence CGM kinematics \citep{Butsky2022} and provide non-thermal feedback, producing outflows with a colder component \citep{Liang2016, Girichidis2018, Farcy2022}. In simulations of Milky Way mass galaxies, CRs have been shown to greatly increase \HI\ and \OVI\ column densities \citep{Ji2020, Salem2016, Kimm2025}. In some cases, CRs can also additionally increase column densities of warm gas tracers such as \CIV\ or C$\,\textsc{iii}$ \citep{Liang2016, Butsky2018}. However, the quantitative impact from CRs on all these observables may depend on galaxy mass \citep{Ji2020}, on the parametrisation of CR feedback \citep{DeFelippis2024} and remains poorly constrained. CRs and AGNs thus presumptively play a major role in the state of the CGM, although their effects and modelling remain complex and warrant further exploration.

\section{Conclusions} \label{sec:conclusions}
    In this work, we analyse three cosmological zoom-in simulations of a single galaxy using the radiative hydrodynamical code RAMSES-RT. The galaxy resides in a halo that reaches $10^{11.6}\rm\,M_\odot$ at $z=1$, and we evolve it using three distinct subgrid models for star formation and feedback: \KI\ \citep{Kimm2017}, \KR\ \citep{Kretschmer2020} and a variation of \KI\ with supernova feedback modelled through delayed cooling \citep{Teyssier2013}, \DC. To isolate the impact from subgrid physics on the CGM, we adjust each model to yield a consistent final stellar mass ($10^{10.2}-10^{10.5}\rm\,M_\odot$ at $z=1$), thereby removing significant discrepancies caused by differences in galaxy masses. Using KROME, we post-process the simulations to compute ionisation fractions of \MgII, \CIV, and \OVI\ to probe the cold, warm, and hot gas phases, respectively.
    Finally, we use RASCAS to generate synthetic lines of sight and estimate column densities for these ions and \HI. Our results can be summarised as follows.

    \begin{itemize}
        \item[$\bullet$] \textbf{The presence of cold and warm ions in the CGM is driven by distinct mechanisms in the different simulations.}\\
            In \KI\ and \KR, most of the \HI\ and \MgII\ is brought by infalling galaxies, leaving high column density gas trails. In \DC, most of the \MgII\ is in the form of elongated filaments stemming from the central galaxy, tracing gas ejected from the ISM. \DC\ also significantly increases column densities by creating low-density channels through which ionising radiation can escape and briefly photoionize the CGM. Additionally, these supernova explosions periodically strip significant amounts of gas from the galaxy, further enriching the CGM. We find similar conclusions for \CIV, except that this gas forms an envelope around structures traced by \MgII.
        \item[$\bullet$] \textbf{The hot CGM can exist in both diffuse and filamentary components depending on the model.}\\
            \OVI\ quickly appears at early times following the first supernova explosions. \OVI\ is then constantly replenished by outflows from the central galaxy as well as from satellites, and kept ionised by high CGM temperatures. \KI\ and \KR\ exhibit both a filamentary \OVI\ structure from infalling satellites and a diffuse component that permeates the whole CGM, driven by feedback. In the \KI\ model, the diffuse component appears to be the most important, while satellite trails play an equally significant role in \KR. In \DC, \OVI\ is fully volume-filling, and powerful feedback largely drives the observed column densities (Fig.~\ref{fig:mapsions}).

        \item[$\bullet$] \textbf{Different subgrid models can be discriminated through quasar absorption line measurements that probe both metals and hydrogen.}\\
            The higher metal mass (resp. gas mass) in \DC\ compared to \KI\ and \KR\ results in higher column densities and covering fractions for \MgII, \CIV, and \OVI\ (resp. \HI, Fig.~\ref{fig:rad_col_cold}-\ref{fig:rad_col_hot}). Additionally, \HI, \MgII, \CIV, and \OVI\ each trace distinct phases of the gas, namely $10^2\rm\,K$, $10^2\rm\,K$, $10^5\rm\,K$, and $10^{5.5}\rm\,K$ for all models (Fig.~\ref{fig:PD_ions},\ref{fig:PD_ions_DC}). Their complementary nature makes these ions extremely powerful in constraining simulations.

        \item[$\bullet$] \textbf{Comparing column densities from observations and simulations is still challenging.}\\
            Lower-energy ions exhibit much larger scatter than higher-energy ions, highlighting their different spatial distributions. Thus, while our median does not necessarily overlap with the bulk of detected absorptions, our distribution does reproduce observed column densities for all species except for \OVI\ in \KI\ and \KR. Due to low statistics and limited sensitivity leading to a significant fraction of upper limits, column densities can give rise to misleading conclusions.
        \item[$\bullet$] \textbf{Covering fractions of weak absorbers provide more robust constraints for simulations.}\\
            Although sensitive to the chosen density threshold, covering fractions offer a more statistically robust characterisation of the absorber distribution than column densities.
            We also find that while higher column density thresholds for covering fractions offer greater reliability, lower thresholds provide stronger constraints.
            Our simulations better reproduce the strong absorber statistics, which may trace tidal tails rather than feedback processes. It is important to compare weak absorbers to constrain feedback models (Fig.~\ref{fig:frac_cold2}, \ref{fig:rad_col_hot2}).

        \item[$\bullet$] \textbf{All three models fail to reproduce the majority of the CGM constraints considered.}\\
            While \DC\ exhibits higher column densities and covering fractions than \KI\ and \KR, all three models fail to reproduce \HI\ and \MgII\ median column densities and covering fractions.
            In \CIV\ and \OVI, both \KI\ and \KR\ median column densities appear well below observational values. While they are also well below \OVI\ covering fractions, \KI\ is in very good agreement with \CIV\ covering fractions up to $\approx50\rm\,kpc$, while \KR\ is slightly below observations.
            Conversely, while \DC\ seems to be in excellent agreement with \CIV\ and \OVI\ column densities, the covering fractions are overestimated (Fig.~\ref{fig:rad_col_cold}-\ref{fig:rad_col_hot}).

        \item[$\bullet$] \textbf{The mismatch between observation and simulations is likely mainly determined by an incorrect ionisation fraction for cold gas, and by the amount of metals in the CGM for hot gas.}\\
            The \DC\ simulation has a higher amount of gas and metals in the CGM, which directly leads to higher column densities for all ions. The amount of metals is the primary variable driving the difference between simulation and observations in \CIV\ and \OVI. However, the large discrepancy between simulation and observation in \HI, \MgII\ is only secondarily impacted by the mass of gas or the metallicity. Although there is a sufficient amount of Hydrogen and Magnesium in the CGM, these species are mostly ionised to higher states. This suggests that gas which should form cold clouds is instead too hot or too diffuse to self-shield (Fig.~\ref{fig:col_max}). 

        \item[$\bullet$] \textbf{In our simulations,} \CIV\ and \OVI\ originate mostly from the outer CGM and the close IGM.\\
            Most of the gas contributing to the computed \CIV\ (resp. \OVI) column densities comes from the outer CGM, up to $\approx1.8\,R_{200}$ (resp $\approx2.2\,R_{200}$) as they originate in the hot diffuse component of the CGM. While \MgII\ appears to originate from the same location as \CIV, this is only a symptom of the absence of \MgII\ in the CGM. Indeed, there is so little \MgII\ in the CGM that column densities are mostly contributed by regions with densities close to our lower limit, integrated over the line-of-sight (Fig.~\ref{fig:artifdmax}). 
    
    \end{itemize}

    We have seen that quasar absorption lines, whether in the form of column densities or covering fractions, provide a powerful probe in differentiating different feedback modes. The roles of \HI, \MgII, \CIV, and \OVI\ are complementary and challenging to reproduce jointly: \OVI\ is largely volume-filling, \HI\ and \MgII\ preferentially trace filaments and dense structures, and \CIV\ spans the transition between these phases. In all our simulations, we find that the cold phase of the CGM is largely underestimated compared to observations. While the warm phase of the CGM is in fair agreement with observations, the hot phase is more challenging to reproduce, with seemingly too much \OVI\ in the delayed cooling simulation and too little in the other simulations. Even though we highlight that our comparison suffers from some caveats, such as the sensitivity to the minimal equivalent width used in covering fractions, our conclusions hold as the discrepancies are significant. To achieve better agreement with observations, our results suggest that the primary requirements are a modified CGM metal distribution for \CIV\ and \OVI\ and higher ionisation fractions for \HI\ and \MgII. We identify numerical resolution and additional physics as the two fundamental levers for these improvements. While higher resolution is a promising solution in principle, its exact impact on the CGM remains a subject of ongoing debate. These aspects are presumably sensitive to the intimate relation between the star formation recipe and the feedback recipe, which highlights the pivotal role of subgrid models in improving future numerical simulations of galaxy formation. active galactic nuclei and cosmic rays may also lead to improvements in the comparison to observations, warranting further exploration in future simulations.

\section*{Acknowledgements}
    We thank Nicolas Bouche for discussions.
    We gratefully acknowledge support from the CBPsmn (PSMN, Pôle Scientifique de Modélisation Numérique) of the ENS de Lyon for the computing resources. The platform operates the SIDUS solution \citep{Quemener2013} developed by Emmanuel Quemener. MR and TK were supported by the National Research Foundation of Korea (NRF) grant funded by the Korean government (Nos. RS-2022-NR070872 and RS-2025-00516961).  
    TK was also supported by the Yonsei Fellowship, funded by Lee Youn Jae.
    This research has made use of the Astrophysics Data System, funded by NASA under Cooperative Agreement 80NSSC21M00561. We also acknowledge the use of Python \citep{VanRossum1995}, Matplotlib \citep{Hunter2007}, NumPy \citep{Harris2020} and Astropy \citep{AstropyCollaboration2013, AstropyCollaboration2018} for this work.


\section*{Data Availability}
    The data generated and used in this article will be shared upon reasonable request to the corresponding author.


\bibliographystyle{mnras/mnras}   
\bibliography{ARCHITECTS_II} 

\appendix
\section{Selection of observational data} \label{sec:Appendix_obs}
    In this appendix, we detail further the selection of the observations against which we compare our simulations. We remind that we study our galaxies with results stacked over $z=1-1.3$, at a redshift higher than most observations. Over this timeframe, our simulations produce galaxies with stellar masses of $10^{10.2}-10^{10.5}\rm\,M_\odot$ within $0.1\,R_{200}$ and are embedded in a dark matter halo of $\approx 10^{11.6}\rm\,M_\odot$.

    \subsection{H\texorpdfstring{$\,\textsc{i}$}{I}}
        For \HI, we compare our simulated column densities to observations from HST with COS \citep{Johnson2015, Prochaska2017, Wilde2021}. \citet{Prochaska2017} covers the same absorption lines as \citet{Werk2014} but with new measurements and either a different estimate of the gas metallicity or more consistent lower limits. 
        \citet{Prochaska2017}, and \citet{Johnson2015} respectively cover a range of stellar masses of $10^{9.6}-10^{11.5}\rm\,M_\odot$ and $10^{8.4}-10^{11.1}\rm\,M_\odot$. The sample selected from \citet{Wilde2021} covers galaxies with stellar masses within $10^{9.9}-10^{11}\rm\,M_\odot$. All observed column densities are at redshift $z<0.4$. For covering fractions, we compare our results against observations from \citet{Wilde2023}, using the COS and Gemini Mapping the Circumgalactic Medium (CGM$^2$) survey. The galaxies in \citet{Wilde2023} cover a redshift range of $0.003 < z < 0.48$ and mass range, and $10^{10}\,\mathrm{M_\odot}<M_\star<10^{10.5}\,\mathrm{M_\odot}$, but provide only two points in our impact parameter range.  

    \subsection{Mg\texorpdfstring{$\,\textsc{ii}$}{II}}
        For \MgII, we compare our simulations to observations from the MagE spectrograph \citep{Chen2010} and Keck with HIRES \citep{Werk2013}. For the data by \citet{Chen2010}, we use \citet{Hummels2013} who converted equivalent widths into column densities following \citet{Draine2011}. These data cover halo masses of $10^{10.6}-10^{13}\rm\,M_\odot$. \citet{Werk2013} cover a stellar mass range of $10^{9.6}-10^{11.5}\rm\,M_\odot$. Due to scarce observations, the range in mass and luminosity of the surveys we selected is wide and extends over up to three orders of magnitude. As with H$\,\textsc{i}$, all observed column densities are at redshift $z<0.4$.

        We compare our simulated covering fractions to observations from \citet{Lan2020}, \citet{Huang2021}, and \citet{Cherrey2025}. In \citet{Huang2021}, the sample covers redshifts of $0.10 < z < 0.48$, and is split betweem star-forming galaxies with $10^{8.3}\,\mathrm{M_\odot}<M_\star<10^{11.4}\,\mathrm{M_\odot}$, and quiescent galaxies, with $10^{9.3}\,\mathrm{M_\odot}<M_\star<10^{11.6}\,\mathrm{M_\odot}$. The model from \citet{Lan2020} takes into consideration the mass, virial radius, and redshift. As our sample is stacked over $1\rm\,Gyr$, we cover a small range for these three quantities (see Fig.~3 in \citetalias{Rey2025}), and we thus compute both the minimum and maximum expected covering fraction over this range. To compute the minimum (resp. maximum) covering fraction with this model, we use the smallest (resp. biggest) redshift, stellar mass and virial radius amongst the timeframes considered ($1.0<z<1.3$), except for passive galaxies where the relation to stellar mass is reversed. The data from \citet{Cherrey2025} is taken from 66 isolated star-forming galaxies from the MEGAFLOW survey at $0.4 < z < 1.5$, with $M_\star>10^{9}\,\mathrm{M_\odot}$. The model from \citet{Cherrey2025} characterises covering fractions as a function of redshift, stellar mass, and star formation rate. Following the same bracketing approach used for \citet{Lan2020}, we use the extrema of these quantities within our simulated sample to define the expected range of covering fractions. We calculate the simulated star formation rate within the $0.1\,R_{200}$.

    \subsection{C\texorpdfstring{$\,\textsc{iv}$}{IV}}
        For \CIV, we compare our simulated column densities to galaxy-selected observations from \citet{Chen2001} and \citet{Bordoloi2014}, respectively relying on the Hubble Space Telescope (HST), Wide Field and Planetary Camera 2 (WFPC), and COS. As with \citet{Chen2010}, points from \citet{Chen2001} are taken from \citet{Hummels2013}, who converted equivalent widths into column densities following \citet{Draine2011}. These points cover stellar masses of $10^{9.5}-10^{11.5}\rm\,M_\odot$. This is also the only data set with redshifts higher than $z=0.4$, going up to $z=0.8920$. Data from \citet{Bordoloi2014} cover a stellar mass range of $10^{8.2}-10^{10.1}\rm\,M_\odot$, right below the stellar masses of our galaxies.

        We compare our simulated covering fractions in \CIV\ to \citet{Schroetter2021}, where the quasar sightlines come from a subsample of 22 galaxies from the MEGAFLOW survey selected to contain strong \MgII\ absorbers. The \CIV\ detections cover a redshift range of $1.0<z<1.5$ and uses a threshold of $W_\CIV = 0.1\,\text{\AA}$.

    \subsection{O\texorpdfstring{$\,\textsc{vi}$}{VI}}
        For \OVI, we use observational data from \citet{Prochaska2011}, \citet{Tumlinson2011} and \citet{Johnson2015}, all using COS. As in the previous sections, the data included all have a redshift $z<0.4$. The stellar mass ranges of the data cover $10^{9.5}-10^{11.5}\rm\,M_\odot$ \citep{Tumlinson2011}, $10^{8.4}-10^{11.1}\rm\,M_\odot$ \citep{Johnson2015} and luminosities of $0.007-2.6\rm\ L*$ \citep{Prochaska2011}.

        For covering fractions, we compare our results to \citet{Kacprzak2015}, who studied isolated galaxies over $0.08<z<0.67$. They use a column density threshold of $W_\OVI = 0.1\,\text{\AA}$ to compute covering fractions. We also include results from \citet{Tchernyshyov2022} in Fig.~\ref{fig:rad_col_hot2} as they use a column density threshold of $N_\OVI = 10^{14.3}\rm\,cm^{-2}$. Their galaxy mass sample has $10^{10.1}\,\mathrm{M_\odot}<M_\star<10^{10.5}\,\mathrm{M_\odot}$ within $0.12 < z < 0.6$.

\section{GS compilation} \label{sec:Appendix_GS}
    We have little data to compare our simulated column densities to, and this is even more limited in covering fractions. For \HI\ and \CIV\ (resp. \OVI), we only have one source (resp two sources) to compare our covering fractions to. To provide an additional constraint, we build covering fractions by combining our previous sample of galaxy-selected observations. We refer to it as the GS compilation.
    
    \subsection{Method}
        We define this combined covering fraction as the fraction of observations composed of either lower limits or actual detections (“detections” from now on) above the chosen column density threshold. The bins in impact parameters were made to contain at least ten observational points. The horizontal extent of the covering fractions computed is the size of the bins, and their vertical extent corresponds to the 15.9 and 84.1 percentiles, estimated by bootstrapping. To do so, we produce 100 random realisations of a sub-sample of eight data points for each bin and then compute the fraction of detections in each of them. For all ions, this method yields results in excellent agreement with the observed covering fractions (see how the black markers compare to the observations in Figs.~\ref {fig:frac_cold}-\ref{fig:rad_col_hot}), which leads us to believe that this approach is reasonable. 

    \subsection{Threshold selection}
            A minimum threshold is necessary to compute the covering fractions, and it must be carefully chosen. If the value is too low, the covering fraction may lose its significance by overestimating the detection limit in several samples. Conversely, taking a column density threshold too high should similarly impact the observations and the simulations, but would produce less constraining results. We use the same threshold for both the GS compilation and to compute covering fractions from our simulations. We now detail the computation of these thresholds.

        \subsubsection*{\HI}
            For H$\,\textsc{i}$, the lower threshold above which the covering fraction is computed for our simulated data and our compilation of observations is the selection criterion of $N_\HI = 10^{14}\rm\,cm^{-2}$ from \citet{Wilde2023} to which we compare our results. This threshold closely matches the minimal detection of our sample ($N_\HI = 1.18\times 10^{14}\rm\,cm^{-2}$).
            These column densities can be converted into equivalent widths by 
            \begin{equation} 
                N_\mathrm{ion} = 1.136\times10^{14} \frac{W_\mathrm{ion}}{f \lambda^2_{1000}}, \label{eq:WtoN}
            \end{equation} 
            assuming an optically thin regime. With $f_{\mathrm{Ly}\alpha}=0.41641$ the oscillator strength of H$\,\textsc{i}$, and $\lambda_{\mathrm{Ly}\alpha} = 1215.6701\,\text{\AA}$ the corresponding wavelength \citep{Kramida2022}, the column density thresholds mentioned earlier respectively correspond to equivalent widths of $W_\HI = 0.54\,\text{\AA}$ and $W_\HI = 0.64\,\text{\AA}$.
    
        \subsubsection*{\MgII}
            In \MgII, we base our column density threshold on an equivalent width of $W_\MgII = 0.3\,\text{\AA}$, which we convert into a column density by using Eq.~\ref{eq:WtoN}. All observational datasets shown in Fig.~\ref{fig:frac_cold} adopt this threshold, except for \citet{Lan2020}, who use $W_\MgII = 0.4,\text{\AA}$.} With $f_\MgII=0.608$ the oscillator strength for Mg$\,\textsc{ii}$, and $\lambda_\MgII = 2796.352\,\text{\AA}$ the corresponding wavelength \citep{Kramida2022}, we obtain a threshold of $N_\MgII = 10^{12.9}\rm\,cm^{-2}$. We also use equivalent widths of $W_\MgII = 0.1\,\text{\AA}$ and $1\,\text{\AA}$ in Fig.~\ref{fig:frac_cold2}.

        \subsubsection*{\CIV}
            As done for the previous ions, we convert the equivalent width threshold of $0.1\text{\AA}$ used in observations from \citet{Schroetter2021} into a column density threshold. 
            With $f_\CIV=0.190$ and $\lambda_\CIV = 1548.202\,\text{\AA}$ \citep{Kramida2022}, we obtain a column density of $N_\CIV = 2.50\times 10^{13}\rm\,cm^{-2}$. This column density is slightly lower than the minimum of the compilation ($N_\CIV = 2.59\times 10^{13}\rm\,cm^{-2}$). In Fig.~\ref{fig:rad_col_hot2}, we use a column density threshold of $N_\OVI = 10^{14.3}\rm\,cm^{-2}$, as used in \citet{Tchernyshyov2022}.

        \subsubsection*{\OVI}
            Lastly, for \OVI, we use an equivalent width threshold of $0.1\text{\AA}$, as in \citet{Kacprzak2015}. With $f_\OVI=0.133$ and $\lambda_\OVI = 1031.912\,\text{\AA}$ \citep{Kramida2022}, we obtain a column density threshold of $N_\OVI \approx 10^{13.9}\rm\,cm^{-2}$.

    \subsection{Impact of detection limits}
        For all ions and column density thresholds considered in this work, we count the number of upper (resp. lower) limits that lie above (resp. below) the chosen column density threshold. Such points are inherently ambiguous, as their true values may lie on either side of the limit.
        Lower limits below the threshold occur only for \MgII\ when adopting a threshold corresponding to $W_\MgII = 1\,\text{\AA}$. These account for just 5 out of 114 measurements and have a negligible impact on our results.
        Upper limits above the threshold are more common. Using the \HI\ and \OVI\ thresholds from \citet{Wilde2021} and \citet{Tchernyshyov2022}, or $W_\MgII = 0.3\,\text{\AA}$ for \MgII, these points represent fewer than 4\% of the sample. However, when adopting thresholds derived from $W_{\rm min} = 0.1,\text{\AA}$, this fraction increases substantially, reaching 16/114 for \MgII, 52/127 for \CIV, and 9/53 for \OVI. To assess the impact of these ambiguous measurements, we recompute the GS compilation covering fractions by treating upper limits as detections. This brackets the range of plausible outcomes: our primary method provides a lower bound, assuming all limits lie below the threshold, while this alternative approach yields an upper bound. We show these upper estimates as grey vertical dashed lines in the relevant figures; their absence indicates a negligible effect.

        Recomputing the GS compilation in this way does not alter our conclusions for \MgII\ (Fig.~\ref{fig:frac_cold2}) or \OVI\ (Fig.~\ref{fig:rad_col_hot}) when using $W_{\rm ion} = 0.1\,\text{\AA}$. The impact on \CIV, however, is substantial (Fig.~\ref{fig:rad_col_warm}). Treating upper limits as detections increases the inferred GS covering fraction by roughly a factor of two, such that even \DC\ underpredicts the data beyond $\approx 40\rm\,kpc$. At the same time, the resulting GS compilation lies well above the \citet{Schroetter2021} model in the outer CGM, highlighting the limitations of this approach for surveys dominated by upper limits. In principle, adopting a higher column-density threshold would reduce the number of ambiguous measurements and yield stronger constraints. In practice, however, the \citet{Hummels2013} sample contains upper limits above most detections, so increasing the threshold does not improve robustness. Excluding \citet{Hummels2013} and relying solely on \citet{Bordoloi2014} leaves too few measurements to construct more than a single coarse bin across the CGM. Consequently, the \citet{Schroetter2021} model remains our most reliable point of comparison for \CIV.In principle, adopting a higher column-density threshold would reduce the number of ambiguous measurements and yield stronger constraints. In practice, however, the \citet{Hummels2013} sample contains upper limits above most detections, so increasing the threshold does not improve robustness. Excluding \citet{Hummels2013} and relying solely on \citet{Bordoloi2014} leaves too few measurements to construct more than a single coarse bin across the CGM. Consequently, the \citet{Schroetter2021} model remains our most reliable point of comparison for \CIV.

\bsp
\label{lastpage}
\end{document}